\documentclass[twocolumn,showpacs,preprintnumbers,amsmath,amssymb,aps,prb,10pt]{revtex4-1}

\usepackage{graphicx}
\usepackage{subfigure}
\usepackage{bm}

\begin{document}

\title{Lifetimes of Confined Optical Phonons and the Shape of a Raman Peak  \\ in Disordered Nanoparticles:  I. Analytical Treatment}

\author{Oleg I. Utesov$^{1,2}$}
\email{utiosov@gmail.com}

\author{Andrey G. Yashenkin$^{1,2}$}

\author{Sergei V. Koniakhin$^{3,4}$}
\email{kon@mail.ioffe.ru}

\affiliation{$^1$Petersburg Nuclear Physics Institute NRC ``Kurchatov Institute'', Gatchina 188300, Russia}
\affiliation{$^2$Department of Physics, St. Petersburg State University, St.Petersburg 199034, Russia}
\affiliation{$^3$Institute Pascal, PHOTON-N2, University Clermont Auvergne, CNRS, 4 Avenue Blaise Pascal, Aubi\`{e}re Cedex 63178, France}
\affiliation{$^4$St. Petersburg Academic University - Nanotechnology Research and Education Centre of the Russian Academy of Sciences, St.Petersburg 194021, Russia}

\date{\today}

\begin{abstract}
Microscopic description of Raman spectra in nanopowders of nonpolar crystals is accomplished by developing the theory of disorder-induced broadening of optical vibrational eigenmodes. Analytical treatment of this problem is performed, and line shape and width are determined as functions of phonon quantum numbers, nanoparticle shape, size, and the strength of disorder. The results are found to be strongly dependent on either the broadened line is separated or it is overlapped with other lines of the spectrum. Three models of disorder, i.e. weak point-like impurities, weak smooth random potential and strong rare impurities are investigated in details. The possibility to form the phonon-impurity bound state is also studied.
\end{abstract}

\maketitle

\section{Introduction}
\label{SIntro}

The properties of very small particles and their ensembles are a subject of current active scientific investigation,
mostly due to their promising applications in material science \cite{behler2009nanodiamond}, quantum computing \cite{andrich2017long,veldhorst2014addressable,riedel2017deterministic}, chemistry \cite{xia2013nanoparticles,lin2018catalysis}, biology and medicine \cite{faklaris2009photoluminescent,kim2016ultrasmall,walling2009quantum,li2011fabrication}, etc.
Among others, the disordered arrays (powders and water suspensions) of crystalline nanoparticles of nonpolar crystals, both semiconducting and diamond-like ones, attract close attention.

Even before being utilized on certain manner, nanopowders need to be attested and certified. For comprehensive certification of a powder such obvious characteristics as chemical formula and crystallographic structure of the material that form the particles of a powder should be supplemented by geometrical parameters of its constituents such as (i) the mean size of a particle, (ii) the particle size distribution function, (iii) the effective faceting number
(in case of nontrivial particle shape), and (iv) the measure of their elongation (if exists), as well as by some characteristics of (v) nanoparticles
intrinsic disorder, surface morphology and phase composition.

In order to examine the nanopowders, several experimental techniques are utilized. The high-resolution transmission electron microscopy (HRTEM) \cite{pichot2008efficient,dideikin2017rehybridization,stehlik2015size,stehlik2016high,skapas2019hrtem,pandey2008slow,osswald2009phonon}, atomic
force microscopy (AFM) \cite{stehlik2015size,stehlik2016high,trofimuk2018effective}, dynamical light scattering~\cite{pecora2000,koniakhin2015molecular,osawa2008monodisperse,koniakhin2020evidence}, X-ray diffraction \cite{koniakhin2018ultracentrifugation,shenderova2011nitrogen,osswald2009phonon}, and Raman spectroscopy (see, e.g., Ref.~\cite{korepanov2020} and references therein) are among them. The latter one is of prime importance because it provides unique precise and nondestructive tool for optical investigations of collective excitations in nanoparticles. Examining the shape of a Raman peak and its position one could extract a great amount of information about the nanoparticles including some parameters mentioned above~\cite{ourDMM, ourEKFG}.

Indeed, since the momentum in a particle is quantized due to finite size quantization effect, the maximum of Raman peak for nanoparticles is shifted as compared to the bulk material with the shift value increasing for smaller particles. Furthermore, the entire discrete spectrum of vibrational modes for the particle of  given shape is peculiar and specific for this particular shape~\cite{ourDMM, ourEKFG}. This manifests itself in the asymmetry of the Raman peak. One can think of restoring the {\it portrait} of a particle from the analysis of peak shape and position, thus formulating a sort of ``inverse problem''. It makes the Raman data analysis very important and challenging issue.

Recently, we proposed two closely related methods of Raman data evaluation~\cite{ourDMM, ourEKFG} which gave much more detailed information about the parameters of
a powder than all previously used variations of the phonon confinement model (PCM)~\cite{richter1981one,campbell1986effects,zi1997comparison,faraci2006modified,korepanov2017,korepanov2017carbon} and other models\cite{gao2019determination,meilakhs2016new}. One of these methods (DMM-BPM)~\cite{ourDMM} consists of the direct solving of dynamical matrix eigenmode problem~\cite{landauI} for a particle with further evaluation of its Raman spectrum. The latter procedure utilizes the proportionality between polarization and deformation in nonpolar crystals with the use of the bond polarization model~\cite{jorioraman}. The (empirically broadened) individual lines of the spectrum  together constitute the first (and subsequent) Raman peaks~\cite{ourDMM}.

Important observation made in Ref.~\cite{ourDMM} and related to discrete spectra of nanoparticles before broadening concerns their fine structure shared
by all particle shapes investigated. Namely, the spectrum starts from a three-fold degenerate
first line which carries the majority (more than 2/3) of
its spectral weight.
This triple line is well-separated from the rest of the spectrum that begins effectively with the 13-th line because of dividing the eigenmodes onto ``Raman active'' and ``Raman silent''  ones. The Raman silent modes do
\begin{figure}[th]
  \centering
  \includegraphics[width=5.0cm]{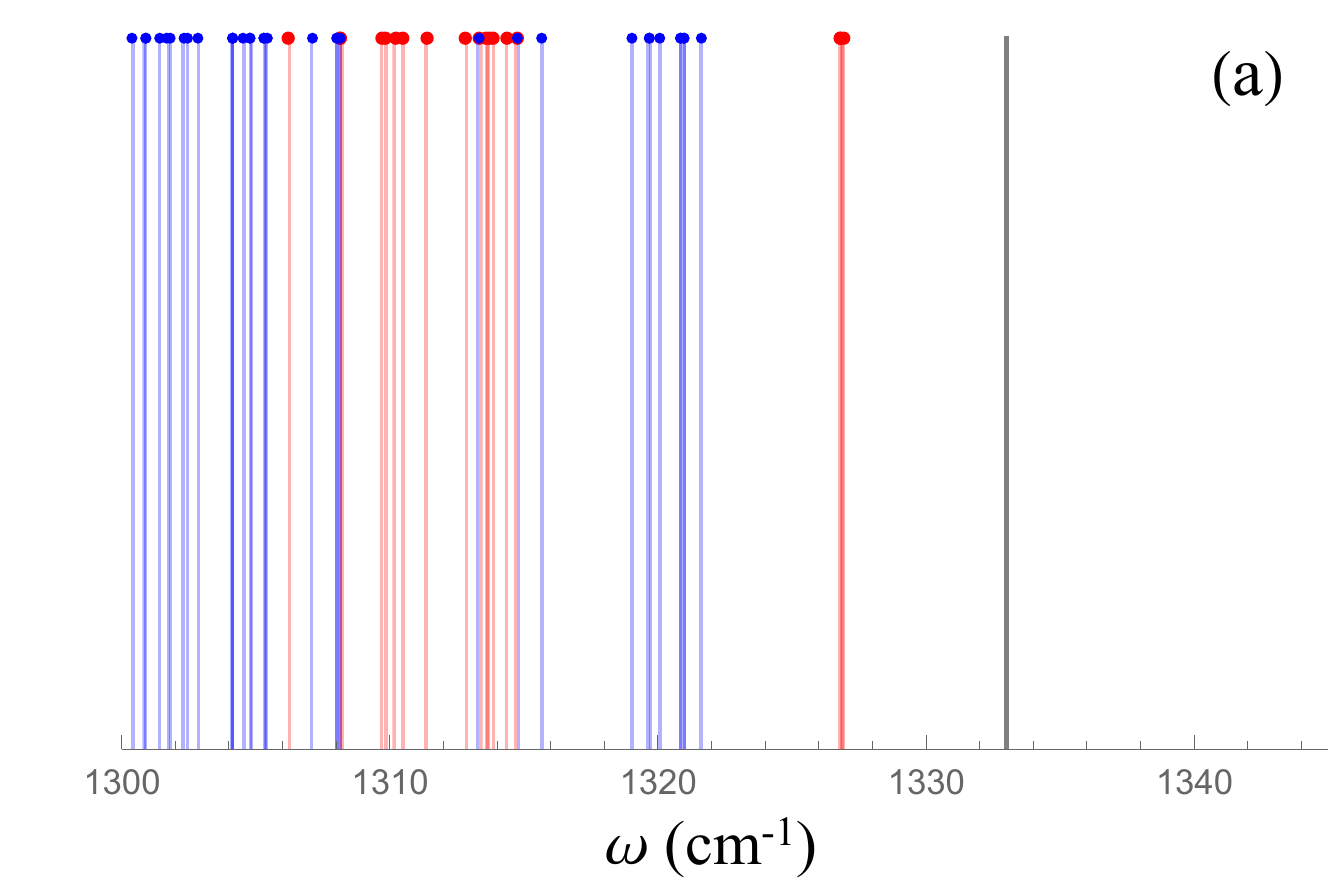}
  \includegraphics[width=5.0cm]{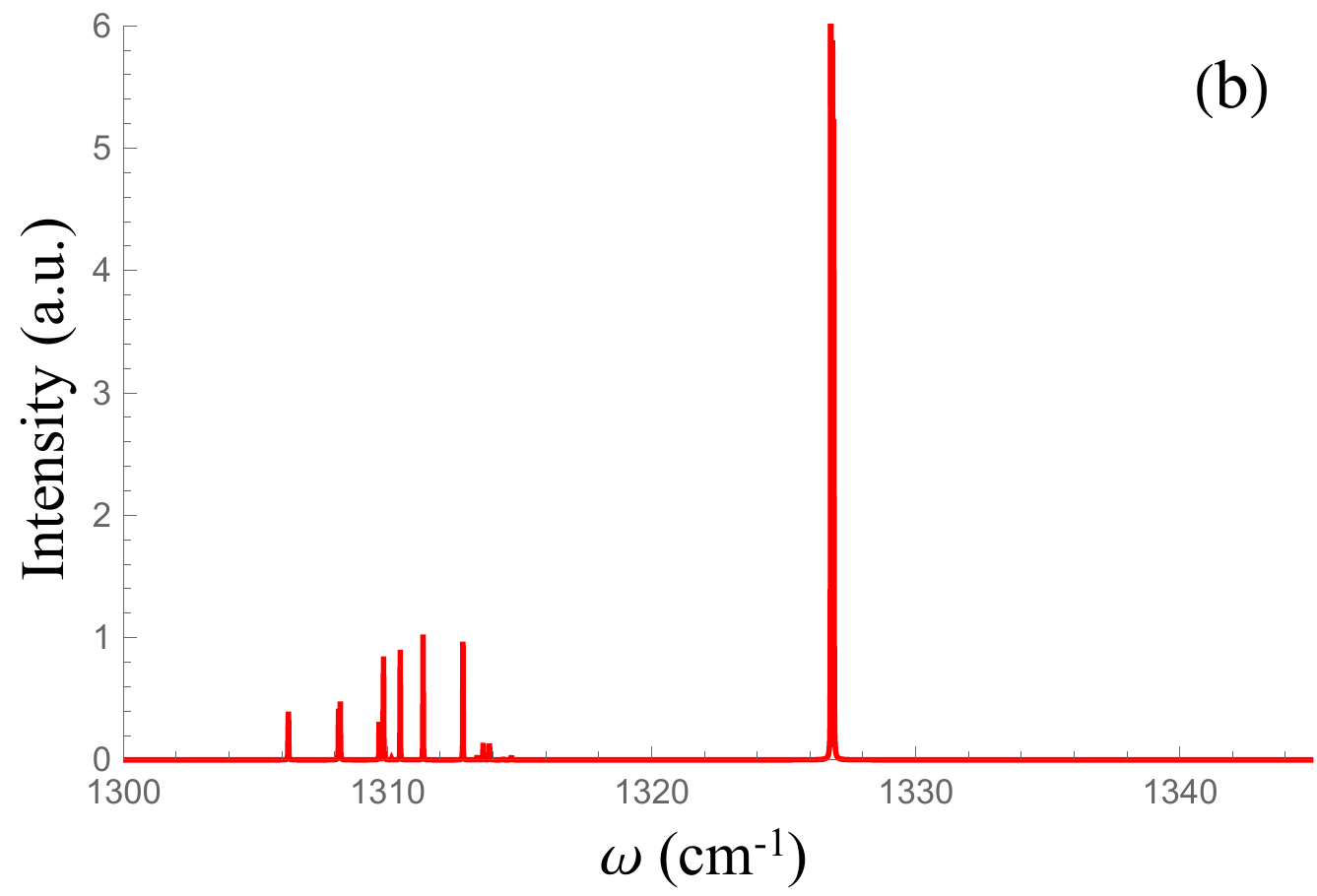}
  \includegraphics[width=5.0cm]{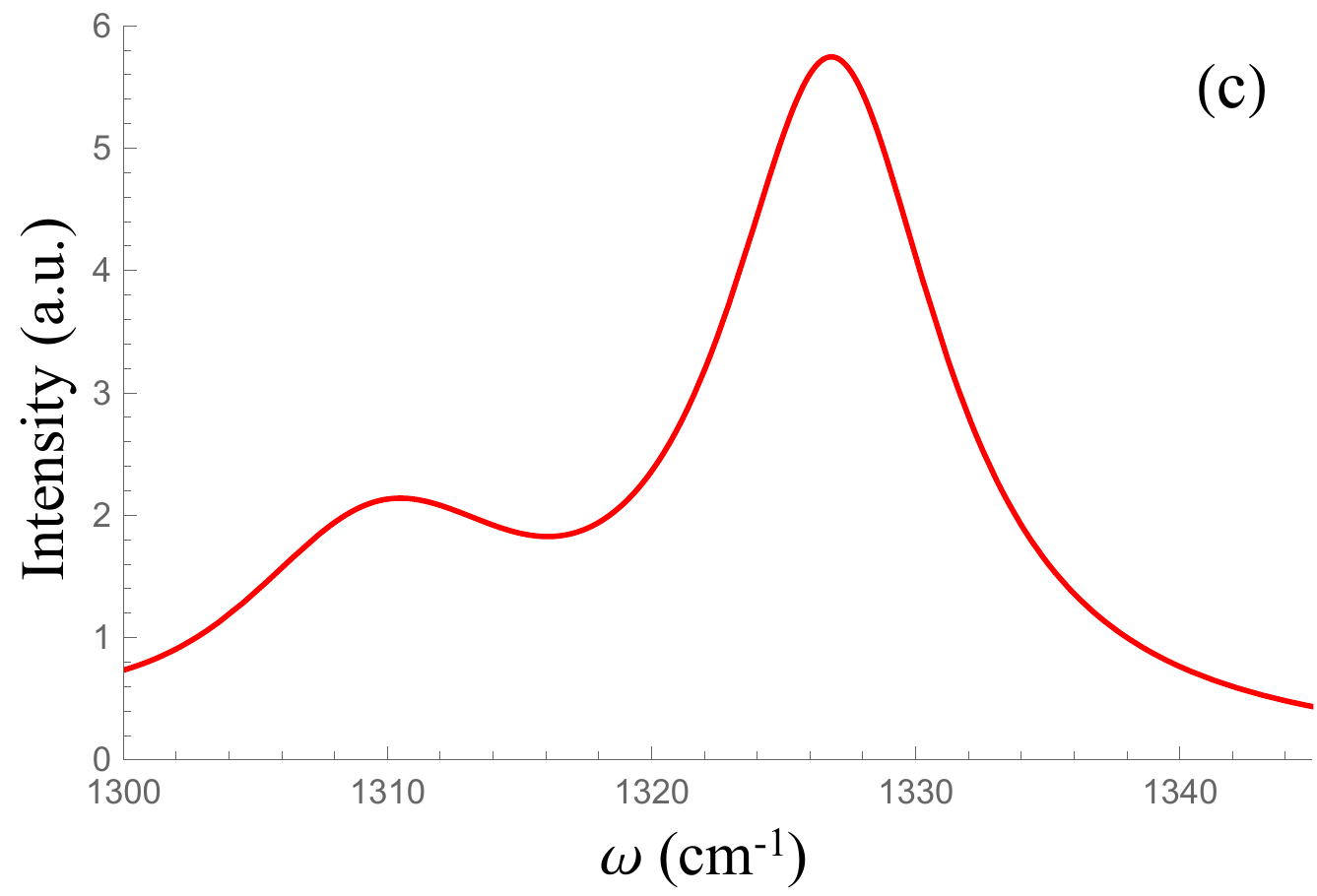}
  \includegraphics[width=5.0cm]{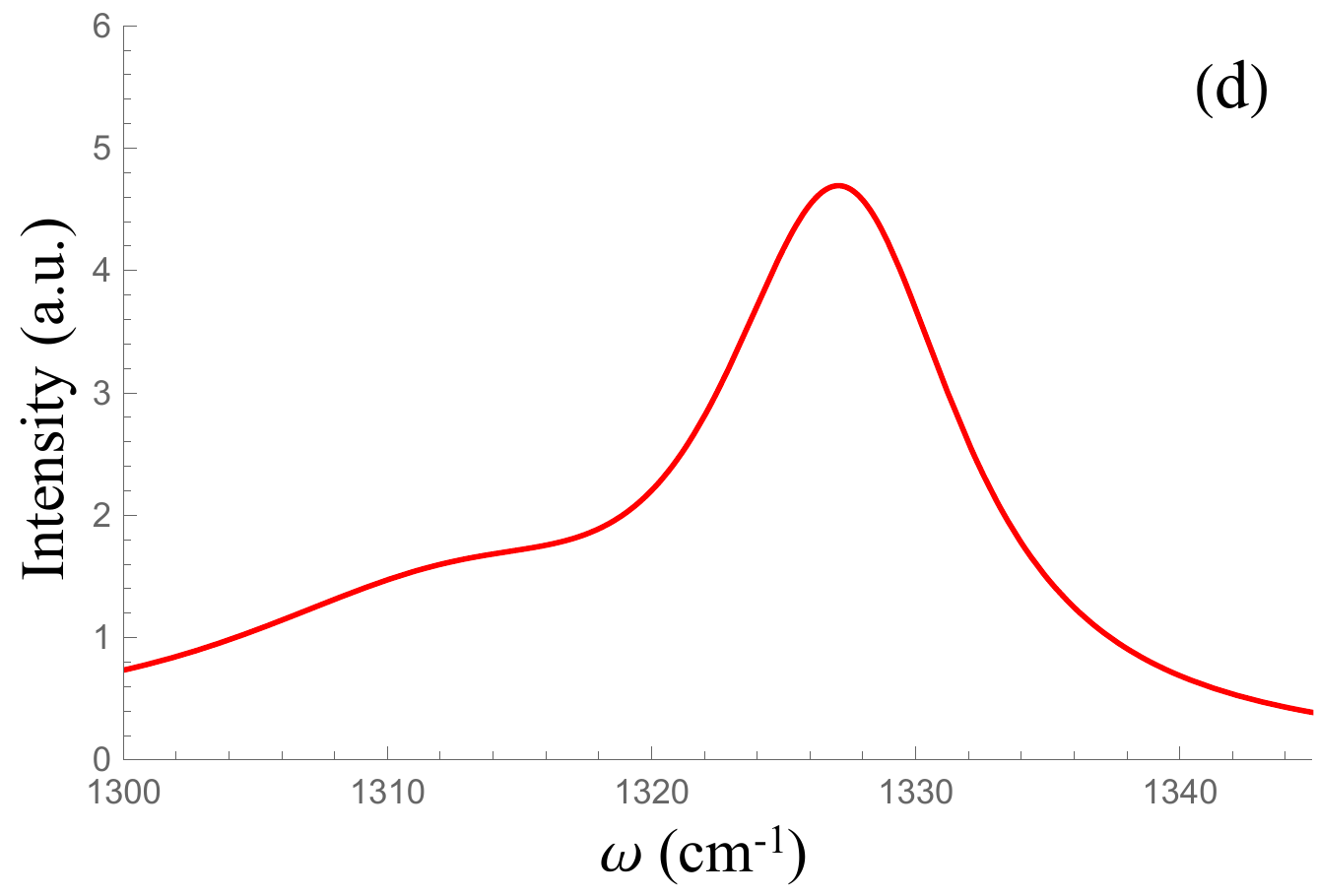}
  \caption{Four steps necessary to obtain the Raman peak for a diamond nanopowder. (a) The phonon spectrum of a 3nm cubic diamond particle consists of Raman-active (red) and Raman-silent (blue) modes. Black vertical line mark the maximal optical phonon frequency $\omega_0$. (b) The Raman spectrum before broadening, only Raman active modes contribute. (c) Phenomenologically broadened Raman spectrum is a smooth function with the main and subsequent peaks. (d) The Raman spectrum of a powder with the mean particle size 3nm and standard deviation 0.3 nm. Only the main Raman peak is clearly seen  while subsequent peaks are smeared out into the slightly sloping left shoulder.
  \label{raman}}
\end{figure}
not contribute to Raman spectra due to symmetry properties of their eigenfunctions. Furthermore, both
Raman silent and Raman active modes form a ``quasi-continuum'' wherein the level spacings
\begin{figure}
  \centering
  \includegraphics[width=6.4cm]{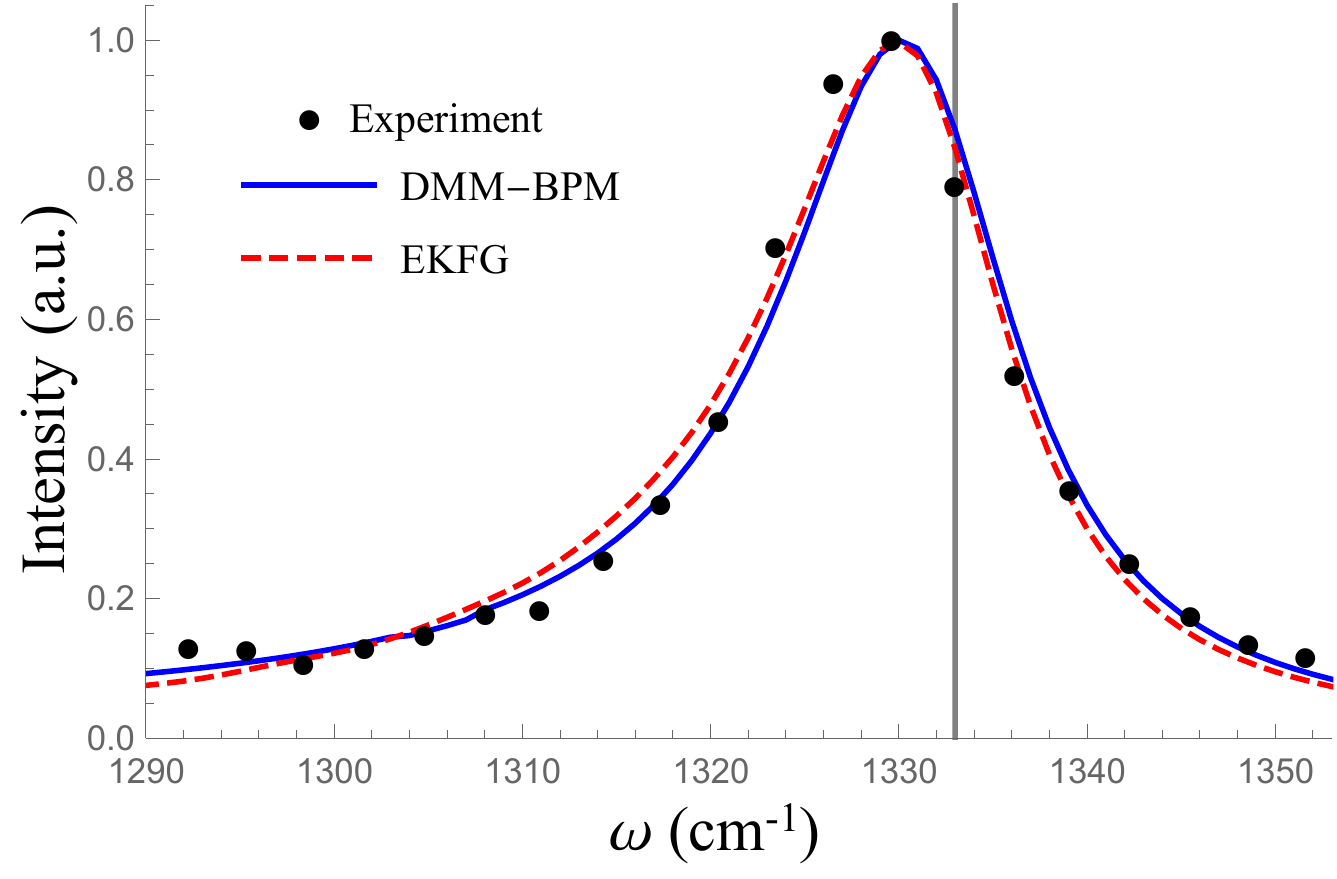}
  \caption{DMM-BPM and EKFG approaches both successfully interpret experimental data. Here solid blue curve (DMM-BPM) and dashed red curve (EKFG) stand for best fits of the experimental Raman spectrum of nanodiamond powder from Ref.~\cite{stehlik2015size} (black dots), the only free parameter is the phonon linewidth $\Gamma$. The vertical grey line represents the maximal optical phonon frequency in the bulk $\omega_0$.
  \label{ramanDMMandEKFG}}
\end{figure}
are essentially smaller than the first gap. This quasi-continuum is split onto several ``bands'' by inter band gaps which do not exceed the first one  (see Fig.~\ref{raman}a).

Terminologically, we shall distinguish between (i) the phonon lines that form the vibrational spectrum of a nanoparticle (see Fig.~\ref{raman}a), (ii) the Raman spectrum which includes only the Raman active vibrational modes with corresponding weights influenced by the photon-phonon scattering matrix elements (see Fig.~\ref{raman}b), and (iii) the main (and possibly subsequent) Raman peak constituted by broadened Raman spectrum lines (see Fig.~\ref{raman}c) of all particle sizes containing in a powder (see Fig.~\ref{raman}d).

Another proposed method (EKFG) replaces original discrete dynamical matrix problem with its long wavelength continuous counterpart which is the Euclidian Klein-Fock-Gordon equation under Dirichlet boundary conditions~\cite{ourEKFG}. Being supplemented by the continuous version of bond polarization model and by the phenomenological line broadening procedure this approach generates Raman spectra almost indistinguishable from those obtained within the DMM-BPM scheme, although {\it before} the line broadening EKFG spectra look a bit oversimplified as compared to the
DMM-BPM ones: they contain one degenerate level in place where the DMM-BPM method yields a bunch of weakly-splitted levels, etc.

It means that the EKFG correctly captures the spectral weight distribution along the energy axis rather than all rigorous details of  the spectrum.
Since the spectral line broadening plays the role of an effective energy averaging, the approximate EKFG method appears to be sufficient for
reproducing all important features of such integral characteristics as the first Raman peak, see Fig.~\ref{ramanDMMandEKFG}.

The imperfection of these theories~\cite{ourDMM, ourEKFG} is the phenomenological character of the line broadening procedure, the linewidth parameter $\Gamma$ is treated as the fitting one with no theoretical analysis of its origin and value. To the best of our knowledge, no detailed theory of Raman peak broadening
in nanoparticles exists in modern literature, all current attempts are having mostly phenomenological or philological character. (One should mention
the paper by Yoshikawa {\it et al.}~\cite{yoshikawa1993raman} who extracted $\Gamma \propto {\rm const} + 1/L$ dependence, with $L$ being the particle size, from the analysis of experimental data.)  The present work aims to accomplish the approach of Refs.~\cite{ourDMM, ourEKFG}, providing us with the microscopic theory of Raman peaks broadening due to nanoparticle {\it disorder and imperfections}, as well as with its numerical verification.

The work consists of two papers, and this is the first one (hereinafter, I). Within the framework of a set of simple models of disorder it theoretically treats the linewidth problem in disordered nanoparticles. Second paper (Ref.~\cite{our4}; hereinafter, II) is devoted to numerical simulations of the same problem, necessary for both verification and justification of analytical results of paper I. In line with paper I numerical paper II allows us to represent the all-round picture of the dirty Raman problem including the effect of realistic disorder.

\begin{figure}
  \centering
  \includegraphics[width=5cm]{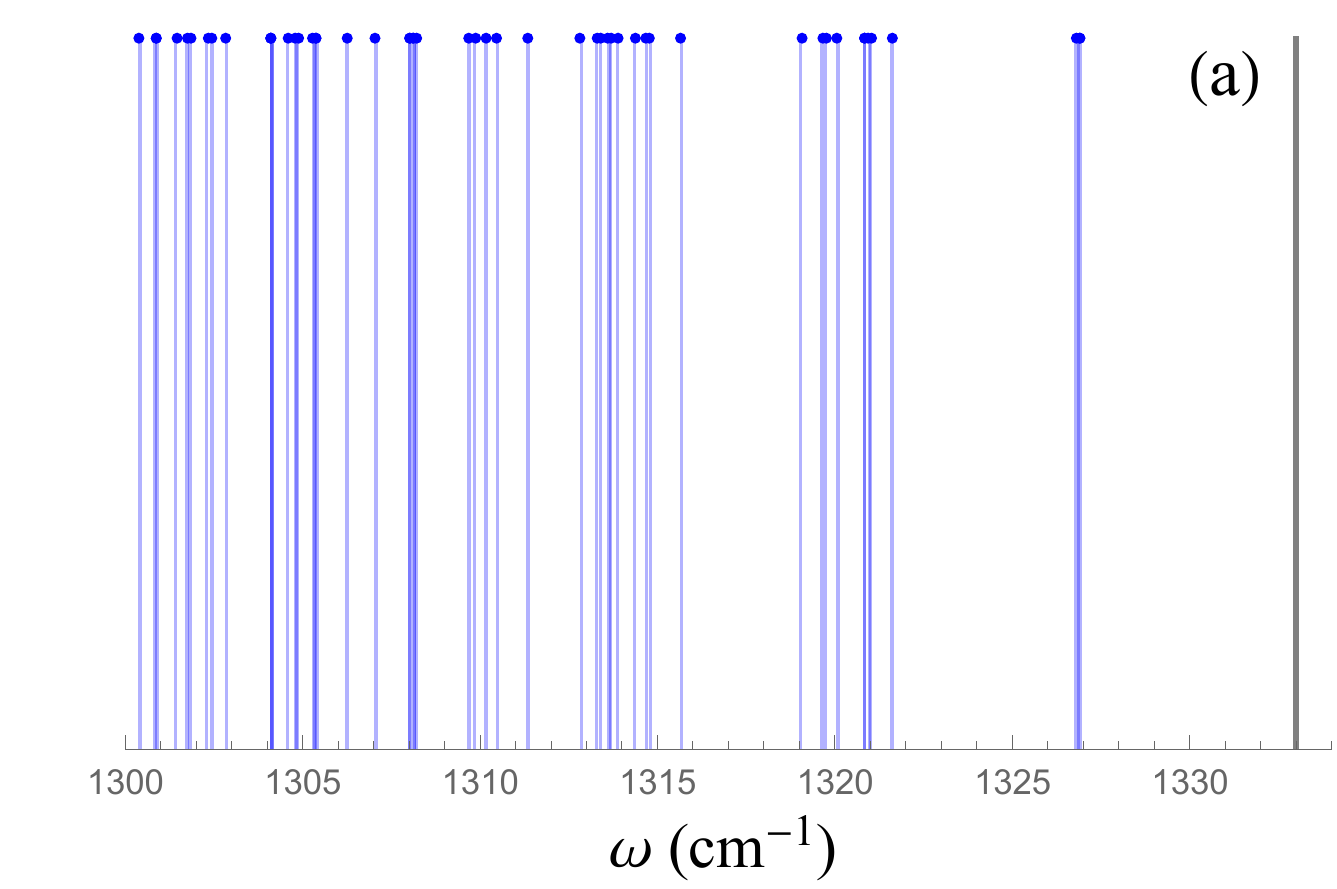}
  \includegraphics[width=5cm]{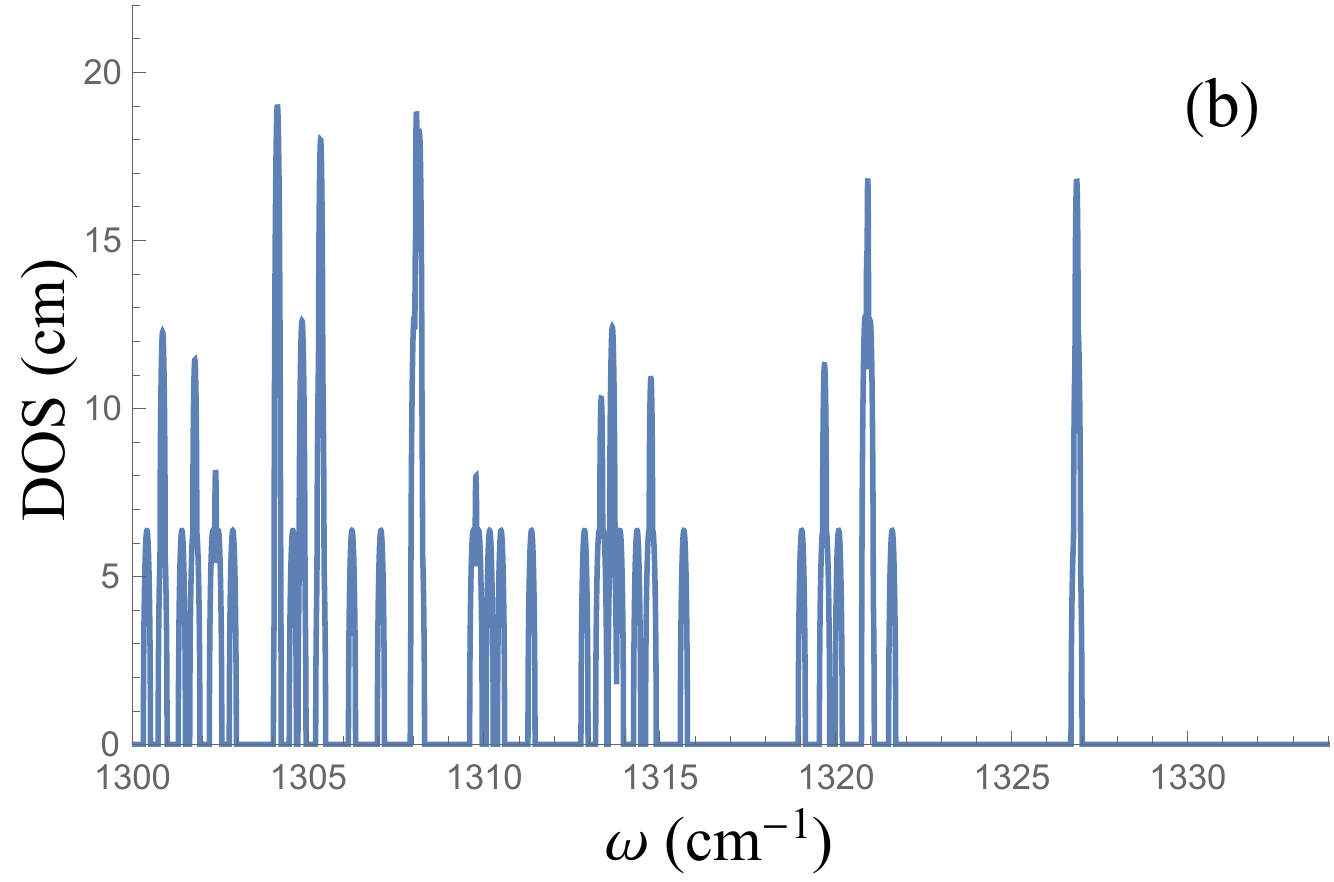}
  \includegraphics[width=5cm]{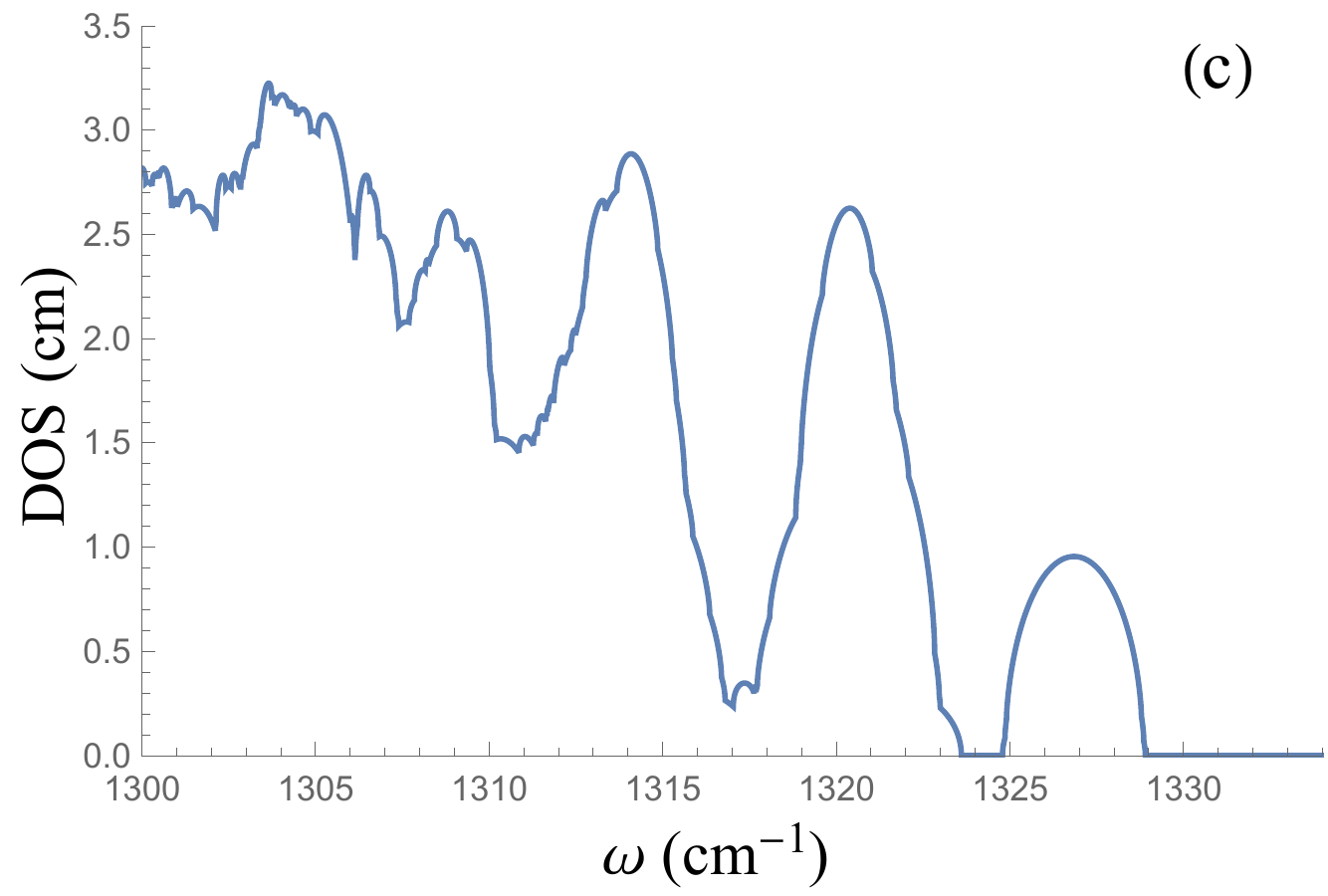}
  \includegraphics[width=5cm]{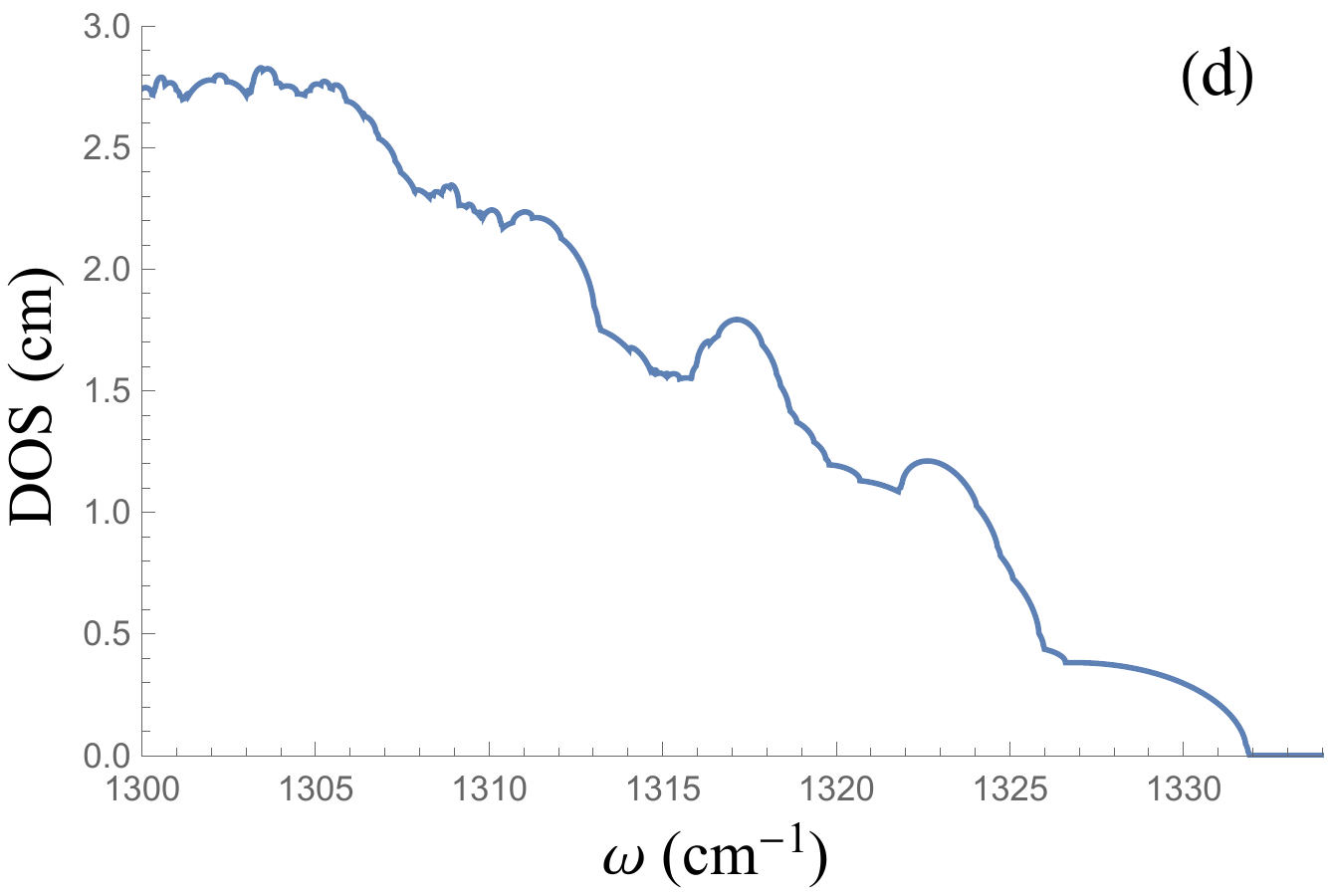}
  \caption{The phonon spectrum of a 3nm cubic nanodiamond calculated with the use of DMM-BPM approach for several amounts of disorder. (a) Clean case. Each dot corresponds to certain eigenfrequency. (b) For weakest disorder the density of states (DOS) consists of discrete peaks. (c) For stronger disorder the lines in quasicontinuum begin to overlap, whereas the first triple  mode is still separated. (d) For strongest disorder all modes are overlapped and the DOS is getting close to the bulk one.
  \label{figbroadening}}
\end{figure}

In order to get preliminary insight into our theory and to understand better the general picture we propose, let us discuss in details Fig.~\ref{figbroadening}. Panel (a) of this Figure represents the typical spectrum of vibrational optical eigenmodes of a particle. As we told above, this spectrum consists of
three-fold degenerate first line (black) and a sequence of ``bands'', the inter level distances between the lines in these bands being essentially smaller than the first gap. Then we introduce disorder into the system. For small amount of disorder it broadens the spectral lines, see Fig.~\ref{figbroadening}b, but the lines remain narrower than the distances between the levels. It is valid for levels lying inside the bands as well as for the first (triple) line. Therefore, we are in the regime of separated levels now for all parts of the spectrum. With disorder increasing, the levels inside the bands start to overlap, while the first line is still separated from the rest of the spectrum, see Fig.~\ref{figbroadening}c. Hence, for this amount of
disorder the band levels crosses over to the continuous regime, while the first triple line remains in the discrete one.  We argue that this ``mixed'' situation is  typical for nanoparticles having the size of order of several nanometers which were investigated in recent experiments~\cite{stehlik2016high}. At last, with further increase of disorder the linewidths become so large that they fill in even the first (largest) gap, making the entire spectrum continuous, see Fig.~\ref{figbroadening}d.

We presented above this detailed picture of the line broadening in order to highlight the importance of the main statement of present work (which,
however, sounds quite evident from general point of view). Namely, we argue that {\it the properties of phonon linewidths in nanoparticles vary dramatically depending on either it is separated level or this level belongs to the continuum}.

More specifically, in paper I we study {\it analytically} the phonon scattering by weak point-like impurities. For disorder induced broadening of spectral lines we observe the linear dependence on the disorder strength parameter $S$ and inverse proportionality to the mean particle size $L$, $\Gamma \propto S / L$, when the broadened levels are overlapped. For separated levels the linewidths behave as $\Gamma \propto \sqrt{S} / L^{3/2}$. At fixed impurity strength there exists a crossover particle size ${\cal L}_{c} \propto a_0 \, S^{-1}$ between these regimes. The strength parameter $S$ is defined as a product of (dimensionless) impurity concentration $c_{imp}$ and (squared normalized) variation of a random parameter of theory (here, the atomic mass $m$), which yields $S = c_{imp} \, \left( \delta m / m \right)^2 $, with $a_0$ being the lattice constant. The prefactors in both regimes are found to be specifically phonon quantum number $n$ and faceting number $p$ dependent quantities, the latter serves for shape parametrization purposes (elongated particles are not considered). We also detect the non-Lorentzian shapes of spectral lines in separated regime of levels broadening and a strong line shape asymmetry in the overlapped regime.

Examining then lowest eigenmodes of phonons subjected to smooth random impurity potential we find that the results crucially depend on the relationship between the mean particle size $L$ and the characteristic spatial scale of a random potential $\sigma$. At $\sigma < L$ the formulas obtained for point-like impurities are applicable in this case, either, the long-range character of the potential generates only small corrections $\sim (\sigma / L)^2$. In the opposite case $L < \sigma$ we get the saturation of $\Gamma (L)$ for separated levels and the linear $L$-dependence of the linewidth for overlapped ones.

We evaluate also the damping of optical phonons in nanoparticles due to strong rare point-like impurities. Investigating separated and overlapped regimes at arbitrary mass variation $|\delta  m| \gtrsim m$ we obtain the same $L$-dependencies of the linewidths as for weakly disordered case, but the disorder strength dependence is more involved. The most interesting situations are the unitary limit for impurity scattering when the mass variation does not enter the formulas, and the regime of resonant scattering when we observe the parametric enhancement of damping. The latter regime is connected with the problem of formation of the impurity-phonon bound state also investigated in present paper.

Technically, our approach in I can be described as follows. In order to address the problem comprehensively, we successively evaluate the broadening of optical phonon spectral lines in nanoparticles due to scattering off  (i) weak point-like Gaussian (Born) impurities and in (ii)  weak {\it smooth} random Gaussian potential, as well as due to (iii) strong dilute binary disorder. When studying the separated levels we apply the self-consistent Born approximation for the first two problems, the diagram technique being formulated in basis of eigenfunctions of the problem. Strong impurities are treated within the T-matrix approximation for both cases. When investigating the overlapped levels we utilize the fact that the eigenfunctions are essentially extended in this case. It allows us to replace them by plane waves thus formulating the ordinary bulk problem within the momentum representation and making the size quantization replacement of the momentum  $q \to q_n (L,p)$ in final formulas. For example, in cubic particles $p=6$ and $q_n = (\pi/L) P_{6} \sqrt{n_x^2 + n_y^2 + n_z^2}$, where $n_{x,y,z}$ are quantum numbers related to size quantization in a cubic box, with $P_{p}$ being the factor converting the linear measure of a particle with faceting number $p$ (e.g., the cube edge) into the diameter of a sphere which contains the same amount of atoms, $P_{6} =(\pi/6)^{-1/3} $.

In paper II we focus on \textit{numerical} studies of disorder induced broadening of vibrational modes in nanoparticles. We use the atomistic DMM and the continuous EKFG model in order to obtain phonon eigenfunctions and corresponding eigenfrequencies. It allows to implement the ``model'' Gaussian disorder for  direct comparison with analytics and to investigate Gaussian and binary disorder, smooth disorder, isotopic impurities, vacancies and NV centers (nitrogen + vacancy) within the unified scheme. This scheme includes numerical evaluation of phonon Green's functions for each state of a pure particle with their subsequent averaging over disorder realizations. The typical statistical sampling was no less than several hundreds. Also, we use continuous EKFG model for the treatment of surface particle imperfections, for which case we introduce and investigate the models of ``peeled apples'' and ``nibbled apples''.

Some results of present work have been reported in Ref.~\cite{ourshort}.

Paper I is organized as follows. Section~\ref{SOverview} contains a short description of approaches of Refs.~\cite{ourDMM, ourEKFG} necessary to make the  understanding of the
line broadening problem more closed and fluent. In Section~\ref{SWeak} we introduce the Hamiltonian and sketch general Green's functions formalism for dirty optical phonons in situations when the energy levels are separated and overlapped. Then we calculate the spectral line broadening within the weak delta-correlated Gaussian impurity potential (point-like Born impurities). In Section~\ref{SSmooth} we extend our treatment onto weak and smooth random potential, while in Section~\ref{SStrong} we elaborate the case of strong rare impurities, the latter being evaluated within the T-matrix method. At last, Section~\ref{SDisc} summarises our analytical results. The comparative discussion of separated and overlapped levels broadening is also presented. Example of application of our general approach (EKFG method for cubic particles) can be found in Appendix.


\section{Brief Overview of DMM-BPM and EKFG Approaches}
\label{SOverview}

This paper is devoted to the analytical treatment of the role of {\it disorder} in the problem of vibrational eigenmodes broadening in nanoparticles, so no detailed knowledge is needed about the theory used to evaluate the ``clean'' spectra. Nevertheless, for the sake of integrity of presentation we supply it with a brief outline of two theoretical methods we advocate for spectral calculations. Their detailed description can be found in Refs.~\cite{ourDMM, ourEKFG}.

The program of interpretation of Raman peaks for a powder of particles of given sort consists of four steps. The {\it first} one is the solution of vibrational eigenproblem providing us with a set of phonon eigenfrequencies and eigenfunctions. In our first approach~\cite{ourDMM} this step is implemented with the use of the {\it dynamical matrix method}~\cite{born1954} which is a direct solution of  $3N \times 3N$ matrix equation of motion for mechanical vibrations:
\begin{equation}\label{DMM}
m  \, \omega^2 r_{l, \alpha} = \sum_{l^\prime=1}^{N} \sum_{\beta =x,y,z} \frac{\partial^2 \Phi}{ \partial r_{l, \alpha} \partial \, r_{l,^\prime \beta}} r_{l^\prime, \beta},
\end{equation}
where $N$ is the number of atoms in a particle, $r_{l, \alpha}$ is the $l$-th atom displacement along direction $\alpha$, $m$ is the mass of the atom,  $\omega$ is the frequency, and $\Phi$ is the total energy of a particle as a function of atomic displacements. The function $\Phi$ could be extracted from any mechanistic theory of crystals; we use the Keating model~\cite{keating1966effect}. This straightforward approach has the only disadvantage: it takes quite a long time to evaluate numerically $3N \times 3N$ matrixes, so the particles exceeding 6 ${\rm nm}$ are hardly analyzable on present day PC clusters.

One of the ways to avoid this difficulty is realized in our second approach. It has been demonstrated in Ref.~\cite{ourEKFG} that the long wavelength limit of discrete DMM problem Eq.~\eqref{DMM} for optical phonons is governed by the continuous Klein-Fock-Gordon equation in the Euclidean space (EKFG) with Dirichlet boundary conditions:
\begin{equation}\label{EKFG}
(\partial^2_t + C_1 \Delta + C_2 ) \, Y = 0, \quad Y|_{\partial \Omega} = 0.
\end{equation}
Here $C_{1,2}$ are the positive constants which can be expressed via the parameters of microscopic theory, $Y$ are the eigenfunctions to be obtained in the course of solution of the problem, $\partial \, \Omega$ is the nanoparticle boundary. This latter equation is much easier to solve for arbitrary particle shape even for the particle size essentially larger than 6 ${\rm nm}$  with the use of routine {\it Mathematica}~\cite{mathem} apparatus.

The {\it second} step of calculations is the elaboration of spectral characteristics obtained at the first step in order to figure out how they manifests themselves in the optical experiment, i.e., as a result of certain photon-phonon interaction. For this aim we used the {\it bond polarization model} which is known to be appropriate for the description of photon scattering in nonpolar crystals under the Raman experimental conditions~\cite{jorioraman}. In this model the polarization tensor $P_{\alpha \beta}$ for $n$-th phonon mode is given by
\begin{equation}\label{BPM}
P_{\alpha \beta} (n) = \sum_{l=1}^{N} \sum_{\gamma} \, M_{l, \alpha, \beta, \gamma} \, r_{l, \gamma} (n),
\end{equation}
with $M_{l, \alpha, \beta, \gamma}$ being some combinations of atomic radius vectors and material constants, which could be expressed via the microscopic parameters of the theory~\cite{ourDMM}, as well.

Analyzing theoretically the spectral properties of diamond nanocrystals of various shapes, we observe a very helpful common feature of them, namely, existence of Raman ``active''
(strongly contributing) and ``silent'' (almost not contributing) modes in the spectrum. This allows us to formulate the approximate analytical version
of the DMM-BPM~\cite{ourDMM} which is much easier to evaluate thus making it applicable for larger particles than the regular DMM-BPM allows. Furthermore, for EKFG method we developed the continuous version of the BPM, and the Raman intensity of $n$-th mode now reads:
\begin{equation}\label{BPME}
  I_n=\left| \int  Y_n \, \text{d}V \right|^2.
\end{equation}
This makes possible to obtain the outcome of EKFG calculations with nearly the same accuracy as it can be done with the use of more direct DMM-BPM method.

As a result of first two steps the Raman spectra are given by dense series of zero-width spectral lines with intrinsic structure peculiar and characteristic for shape and sort of particles studied. The ${\it third}$ step needed to describe the Raman experiment is to broaden these lines replacing zero-width delta functions by Lorentzians and thus introducing the damping of individual eigenmodes. This damping is known to be much larger for nanoparticles than for relevant bulk materials~\cite{ager1991}. In our previous calculations, we treated the linewidth $\Gamma$ as a fitting parameter. The present work is intended to certify disorder as the main microscopic source of the phonon line broadening observed in Raman experiments and to express fitting parameters via the microscopic characteristics of disordered nanocrystals.

After the spectral lines are properly broadened, we end up with the Raman peak (or peaks) as it would exist for a powder of equal sized particles. The {\it fourth} step of calculations is to include the size distribution function. The easiest way to do it is to apply for spectral lines  scaling arguments developed in~\cite{ourEKFG} within the EKFG approach:
\begin{equation}\label{scaling}
I_{L_2} \, (\omega) = \left( \frac{L_2}{L_1} \right)^3 \, I_{L_1} \!\! \left( \omega_0 - (\omega_0 - \omega) \, \left( \frac{L_2}{L_1} \right)^2 \right).
\end{equation}
Here $I_{L_{1,2}} (\omega)$ are the spectra of particles both having the same shape but different sizes $L_1$ and $L_2$, respectively.

Empirically, the EKFG scaling Eq.~\eqref{scaling} may be extended onto DMM-BPM approach, as well, which can be justified by the similarity of spectra obtained with the use of both these methods. It allows to incorporate the size distribution into the theory without boring recalculation of spectra for each particle size containing in distribution function, as one should do at the first glance.

Notice that the  Raman peaks calculated within the DMM-BPM and EKFG methods look very similar. They also fit existing experimental data for small nanoparticles much better than previous theories even with empirical broadening procedure undertaken instead of the third step of this Section approach (see~\cite{ourDMM,ourEKFG} and Fig.~\ref{ramanDMMandEKFG}).

\section{Weak point-like impurities}
\label{SWeak}

In this Section we introduce the diagram technique for the disordered phonon problem and discuss the phonon scattering by Born impurities in two qualitatively different regimes of separated and overlapped energy levels.

\subsection{General formalism}
\label{SSGeneral}

Let us start the derivation of the Hamiltonian of ``dirty phonons'' from the more general one:
\begin{equation}\label{hgen0}
  {\cal H} = \sum_l \frac{p^2_l}{2 m_l} +  \frac{1}{2} \, \sum_{l l^{\prime}} K_{ll^{\prime}} \left( \mathbf{r}_l - \mathbf{r}_{l^{\prime}} \right)^2,
\end{equation}
where the first sum runs over all lattice sites $l$, and the atomic mass $m_l$ varies from site to site. The second sum includes all pairs of atoms, with spring rigidities being  $K_{ll^{\prime}}$, and $\mathbf{r}_l$ stand for corresponding displacements.

As we mentioned above the phonon line broadening problem in a nanoparticle should be treated separately depending on whether this broadening results in level overlaps or not. For weak enough disorder and/or small enough particles we are definitely in the regime of separated levels. In order to describe this regime it is convenient to use the Green's functions formalism formulated in the basis of eigenfunctions of a given problem.

In a particle containing $N$ atoms there are $3N$ normalized to unity vibrational modes $\mathbf{Y}_n(\mathbf{R}_l)$ with energies $\omega_n$. Here $n$ is the generalized (multicomponent) quantum number of our eigenproblem. Assuming all masses to be equal to each other (as, e.g., in diamond) for atom displacements and momenta we have
\begin{equation}\label{rlQ}
  \mathbf{r}_l = \frac{1}{\sqrt{2 m}} \sum_n \frac{\mathbf{Y}_n(\mathbf{R}_l)}{\sqrt{\omega_n}} (b_n + b^\dagger_n)
\end{equation}
and
\begin{equation}\label{plQ}
  \mathbf{p}_l = \frac{i \sqrt{m}}{\sqrt{2}} \sum_n \mathbf{Y}_n(\mathbf{R}_l) \sqrt{\omega_n} (b^\dagger_n - b_n),
\end{equation}
respectively. Using Eqs.~\eqref{rlQ} and~\eqref{plQ} one gets the Hamiltonian in the form ${\cal H}={\cal H}_{ph} +{\cal H}_{imp}$, where the first term yields the phonon energy
\begin{equation}\label{h0}
{\cal H}_{ph} = \sum_{n} \, \omega_{n}  \left( \, b^{\dagger}_n b_n + 1/2 \right).
\end{equation}
The second term ${\cal H}_{imp}$ specifies how the disorder affects vibrational modes. Hereinafter we shall assume for simplicity that disorder appears in the problem via the mass variation only (another source of disorder would be random rigidities). For simplest point-like impurities the perturbation of bare Hamiltonian $\mathcal{H}_{ph}$ reads:
\begin{equation}\label{h imp}
{\cal H}_{imp} = \frac{1}{2} \sum_{l} \delta m^{-1}_l p_{l}^2,
\end{equation}
with the inverse mass variation $\delta m^{-1}_{l}$ given by
\begin{equation}\label{delta m}
\delta m^{-1}_l = \frac{1}{m +  \delta m_l} - \frac{1}{m} \approx - \frac{\delta m_l}{m^2} ,
\end{equation}
where
\begin{equation}\label{m m-}
  m  =  \langle \, m_l \, \rangle   ,
\end{equation}
and the random masses are supposed to be Gaussian delta-correlated quantities with zero averages
\begin{equation}\label{delta m av}
\langle \, \delta m^{-1}_l \, \rangle =0, \quad \langle \, \delta m_l \, \rangle = 0,
\end{equation}
and delta-functional pairwise correlators
\begin{equation}\label{delta m 2 av}
\langle \, \delta m_l \, \delta m_{l^{\prime}} \, \rangle /m^{2} \, = \,
\delta_{l l^{\prime}} \,  c_{imp} \, \left( \frac{\delta m}{m} \right)^2 \, \equiv \, \delta_{l l^{\prime}} \,  S .
\end{equation}
Here $S \ll 1$ is the dimensionless strength of disorder. ``Strong'' impurities ($S \lesssim 1$) with nonzero average will be considered in Section~\ref{SStrong} of this paper.

\begin{figure}
  \centering
  \includegraphics[width=6.4cm]{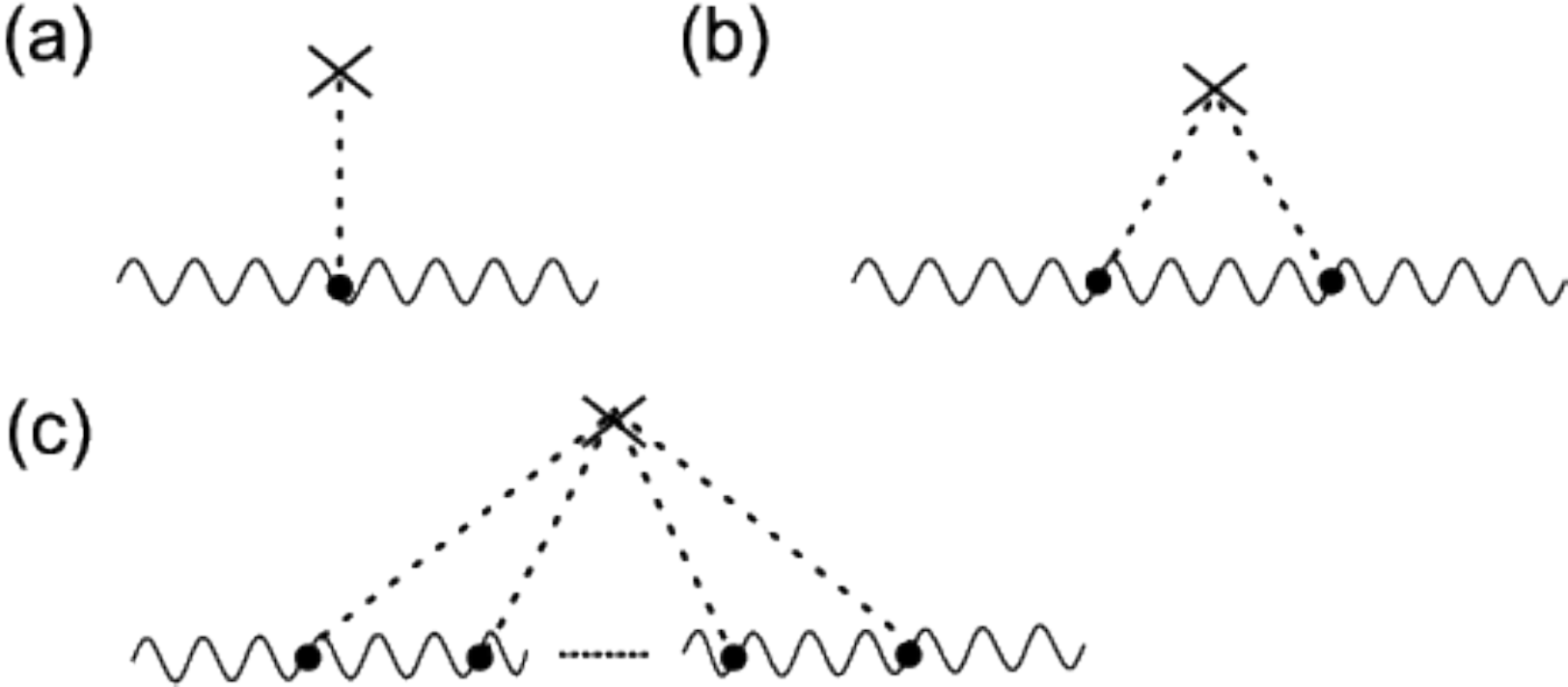}
  \caption{\label{Diagrams} Diagrams for phonon scattering by disorder evaluated in present paper. (a) Single scattering by the impurity does not lead to phonon damping. (b) Diagram giving the leading contribution to the phonon self-energy in the Born regime. The self consistent Born approximation appears when replacing the bare phonon propagator under the ``impurity arch'' by the full one.  For strong rare impurities one should sum up all diagrams of the type (c) taking into account multiple scattering by the same impurity (T-matrix approximation).
\label{DiagScat}}
\end{figure}

Using Eqs.~\eqref{rlQ} and~\eqref{plQ} we can write the impurity-induced perturbation in the form
\begin{eqnarray}\label{h imp2}
  {\cal H}_{imp} =  - \frac{m}{4} \sum_{l, n, n^\prime}  \, \mathbf{Y}_n(\mathbf{R}_l) \cdot \mathbf{Y}_{n^\prime}(\mathbf{R}_l)
  \, \delta m^{-1}_l \\ \sqrt{\omega_n \omega_{n^\prime}} (b^\dagger_n-b_n)(b^\dagger_{n^\prime}-b_{n^\prime}). \nonumber
\end{eqnarray}
Let us introduce the Green's function $- i \langle \hat{{\rm T}} \, \phi_n \phi_{n} \rangle$ for operators  $\phi_n = i(b^\dagger_n - b_n)$, where $\hat{{\rm T}}$ stands for T-ordering. Upon averaging over impurity configurations, the self-energy $\Pi_n (\omega)$ arising due to phonon scattering by disorder enters this Green's function on the following way:
\begin{equation}\label{DQ}
D_n(\omega)= \frac{2  \omega_n}{\omega^2 - \omega^2_n - 2 \omega_n \Pi_n (\omega)}.
\end{equation}
To the leading order in the impurity strength $S$ only the diagram shown in Fig.~\ref{DiagScat}b contributes to the phonon self-energy (diagram Fig.~\ref{DiagScat}a is zero due to condition Eq.~\eqref{delta m av}), and corresponding contribution reads:
\begin{equation}\label{PiQ}
\Pi_n (\omega) = \frac{S \omega_n}{16} \sum_{l, n^\prime}  \, \left[ \mathbf{Y}_n(\mathbf{R}_l) \cdot \mathbf{Y}_{n^\prime}(\mathbf{R}_l) \right]^2 \, \omega_{n^\prime} D_{n^\prime} (\omega).
\end{equation}
This equation will be solved in the next Subsection.

When disorder increases the levels start to overlap. Another way to obtain overlapped states is to treat large particles. For both these situations it is convenient to use the ordinary (bulk) diagram technique in the momentum space and then incorporate the finite size quantization  $q \rightarrow q_n(L, p)$ in final formulas,
the quantization rule $q_n(L, p)$ depends on the shape of a particle.

For bulk crystals one can use standard expressions for atomic displacements and momenta via phonon creation-annihilation operators $b^{\dagger}_{{\bf q} \nu} (b_{{\bf q} \nu})$:
\begin{equation}\label{rl}
{\bf r}_l  = \frac{1}{\sqrt{2N m}} \sum_{{\bf q} \nu} \frac{{\bf P}_{{\bf q}}}{\sqrt{\omega_{{\bf q} \nu}}} \left( \, b^{\dagger}_{{\bf q} \nu} e^{- i {\bf q} {\bf R}_l} + b_{{\bf q} \nu} e^{i {\bf q} {\bf R}_l} \, \right) ,
\end{equation}
\begin{equation}\label{pl}
{\bf p}_l  = \frac{i\sqrt{m}}{\sqrt{2N}} \sum_{{\bf q} \nu} {\bf P}_{{\bf q}}\sqrt{\omega_{{\bf q} \nu}} \left( \, b^{\dagger}_{{\bf q} \nu} e^{- i {\bf q} {\bf R}_l} - b_{{\bf q} \nu} e^{i {\bf q} {\bf R}_l} \, \right).
\end{equation}
Here ${\bf P}_{\bf q}$ is the phonon polarization and index $\nu$ runs over all branches of the optical spectrum. For longitudinal phonons ${\bf P}_{\bf q}={\bf e}_{\bf q}={\bf q}/q$, and for transverse ones ${{\bf P}_{\bf q} \perp {\bf e}_{\bf q}}$. Using Eqs.~\eqref{rl} and~\eqref{pl} we obtain the Hamiltonian ${\cal H}_{ph}$ in the form
\begin{equation}\label{h0}
{\cal H}_{ph} = \sum_{{\bf q} \nu} \, \omega_{{\bf q} \nu}  \left( \, b^{\dagger}_{{\bf q} \nu} b_{{\bf q} \nu} + 1/2 \right) ,
\end{equation}
where $\omega_{{\bf q} \nu}$ is the bare phonon frequency. For the sake of simplicity, the phonon spectrum is taken to be identical in all  branches; generalization onto the case of different parameters is straightforward. We shall be interested in optical phonons near the Brillouin zone center,
where their dispersion is well approximated by the formula
\begin{equation}\label{omega 0}
\omega_{\bf q} = \omega_0 - \alpha q^2 = \omega_0\left[1 - F(q a_0)^2\right].
\end{equation}
Here we introduce the dimensionless parameter of flatness of the optical phonon spectrum $F$:
\begin{equation}\label{flat}
  F  = \frac{1}{2 \omega_0 a_0^2} \left( \frac{\partial^2 \omega_\mathbf{q}}{\partial q^2} \right)_{\mathbf{q}=0},
\end{equation}
and $a_0$ is the lattice parameter. Relations between phenomenological parameters $\omega_0$ and $\alpha$ and the parameters of microscopic Keating model~\cite{keating1966effect} for materials under investigation can be found, for instance, in Refs.~\cite{ourDMM} and \cite{ourEKFG}.

We can rewrite the interaction term Eq.~\eqref{h imp} as follows:
\begin{eqnarray}\label{h imp1}
&& {\cal H}_{imp} =  \frac{m}{4 N} \!\!\! \sum_{l, \, {\bf q}_{1,2}, \nu} \!\!\! ({\bf P}_{{\bf q}_1} \cdot {\bf P}_{{\bf q}_2}) \,
  \delta m^{-1}_l \sqrt{\omega_{{\bf q}_1 \nu} \, \omega_{{\bf q}_2 \nu} }   \\
&& \times \left[ b^{\dagger}_{{\bf q}_1 \nu} b_{{\bf q}_2 \nu} e^{i ({\bf q}_2 - {\bf q}_1) {\bf R}_l} - b_{{\bf q}_1 \nu} b_{{\bf q}_2 \nu} e^{i ({\bf q}_2 + {\bf q}_1) {\bf R}_l} \right] + \text{H.C.}, \nonumber
\end{eqnarray}
where H.C. stands for Hermitian conjugate. The diagram technique for the Hamiltonian Eqs.~\eqref{h0} and \eqref{h imp1} (see, e.g., Ref.~\cite{koniakhin2017}) is very similar to the one widely used for disordered electrons (cf. Ref.~\cite{agd}). The Green's
function $D (\omega, {\bf q})$ is built upon the bosonic field operators
$ { \phi ({\bf q}) \propto b_{\bf q} - b^{\dagger}_{-{\bf q}} }$.
The bare phonon propagator is given by
\begin{equation}\label{D_0}
D_0 (\omega, {\bf q})= \frac{2 \omega_{\bf q}}{\omega^2 - \omega^2_{\bf q}  + i 0}
\end{equation}
After averaging over impurity configurations, the disorder-induced self-energy $\Pi_{{\bf q}} (\omega)$ enters this equation on the following way:
\begin{equation}\label{D}
D^{-1} (\omega, {\bf q})= D_{0}^{-1} (\omega, {\bf q}) - \Pi_{\bf q} (\omega).
\end{equation}
To the leading order in $S$ the analog of Eq.~\eqref{PiQ} reads:
\begin{equation}\label{Pi}
\Pi_{\bf q} (\omega) = \frac{S \, \omega_{\bf q}}{16 N} \sum_{{\bf k}}  \, \omega_{\bf k} D_0 (\omega, {\bf k}),
\end{equation}
where the proper choice of coordinates allowed us to eliminate the contribution from one of the phonon branches.

It should be mentioned the similarity between our formulas for separated and overlapped levels, cf. Eqs.~\eqref{DQ} and~\eqref{D} as well as Eqs.~\eqref{PiQ} and~\eqref{Pi}. However, the final results for these two cases are drastically different.

\subsection{Separated levels}
\label{SSWeakSep}

When using the formal perturbation theory in the parameter $S$ the phonon self-energy on the r.h.s. of Eq.~\eqref{PiQ} should be omitted. It can be shown that no broadening of the phonon line appears in this case. We shall use the self-consistent Born approximation keeping the self-energy on the r.h.s. of Eq.~\eqref{PiQ} finite, and thus obtaining $\Pi_n(\omega)$ self-consistently. This method resembles the approach used for disorder-induced broadening of Landau levels in the two-dimensional degenerate electron gas subject to high magnetic fields~\cite{ando1974}. Following the same line we consider equation
\begin{equation}\label{PiQW}
  \Pi_n (\omega) = \frac{S \omega_n}{8} \sum_{l, n^\prime}  \,\frac{  \left[ \mathbf{Y}_n(\mathbf{R}_l) \cdot \mathbf{Y}_{n^\prime}(\mathbf{R}_l) \right]^2 \,  \omega^2_{n^\prime}}{ \omega^2 - \omega^2_{n^\prime} - 2 \omega_{n^\prime} \Pi_{n^\prime}(\omega)}.
\end{equation}
One can easily see that for non-overlapped levels this equation contains effectively only the term with $n^\prime = n$. For non-degenerate levels it yields:
\begin{equation}\label{PiQW1}
  \Pi_n (\omega) = \frac{2 \, c^2_n(p) \, S \, \omega^3_n}{N}  \,\frac{ 1}{ \omega^2 - \omega^2_{n} - 2 \omega_{n} \Pi_{n}(\omega)}.
\end{equation}
Here $c^2_n(p) = N \sum_l \left[ \mathbf{Y}_n(\mathbf{R}_l)\right]^4/16 $ is certain shape $p$ and quantum number $n$ dependent coefficient. Solving Eq.~\eqref{PiQW1} we obtain:
\begin{equation}\label{PiQW2}
  \Pi_n(\omega) = \frac{\displaystyle{\omega^2 - \omega^2_n - \sqrt{\left(\omega^2 - \omega^2_n\right)^2 - \frac{16 c^2_n(p) S \omega^4_n}{N}}}}{\displaystyle{4 \omega_n}}.
\end{equation}
When the square root in Eq.~\eqref{PiQW2} is imaginary it determines the phonon damping which leads to the well-known semi-circle law for density of states~\cite{ando1974}. Similar to the case of disordered electrons in high magnetic fields, the range of validity for this self-consistent solution is limited by  proximities of semi-circles maxima, whereas near their edges one should take into account the phonon scattering by rare configurations of impurities which results in exponential
tails in density of states (see Ref.~\cite{stone1992} and references therein). However, if we are not interested in these details, one can use the crude Lorentzian (on-shell) approximation.  This approximation ($\omega = \omega_n$) in Eq.~\eqref{PiQW2} leads to the phonon damping in the form
\begin{equation}\label{GDMM1}
  \Gamma_n =  \omega_n \, c_n(p) \sqrt{\frac{S}{N}}.
\end{equation}
The definition of $\Gamma$ we use in papers I and II differs from that of Refs.~\cite{ourDMM, ourEKFG, ourshort} where the width of Lorenzian has been parameterized by $\Gamma/2$ rather than $\Gamma$ of present work.

\begin{figure}
  \centering
  \includegraphics[width=6.4cm]{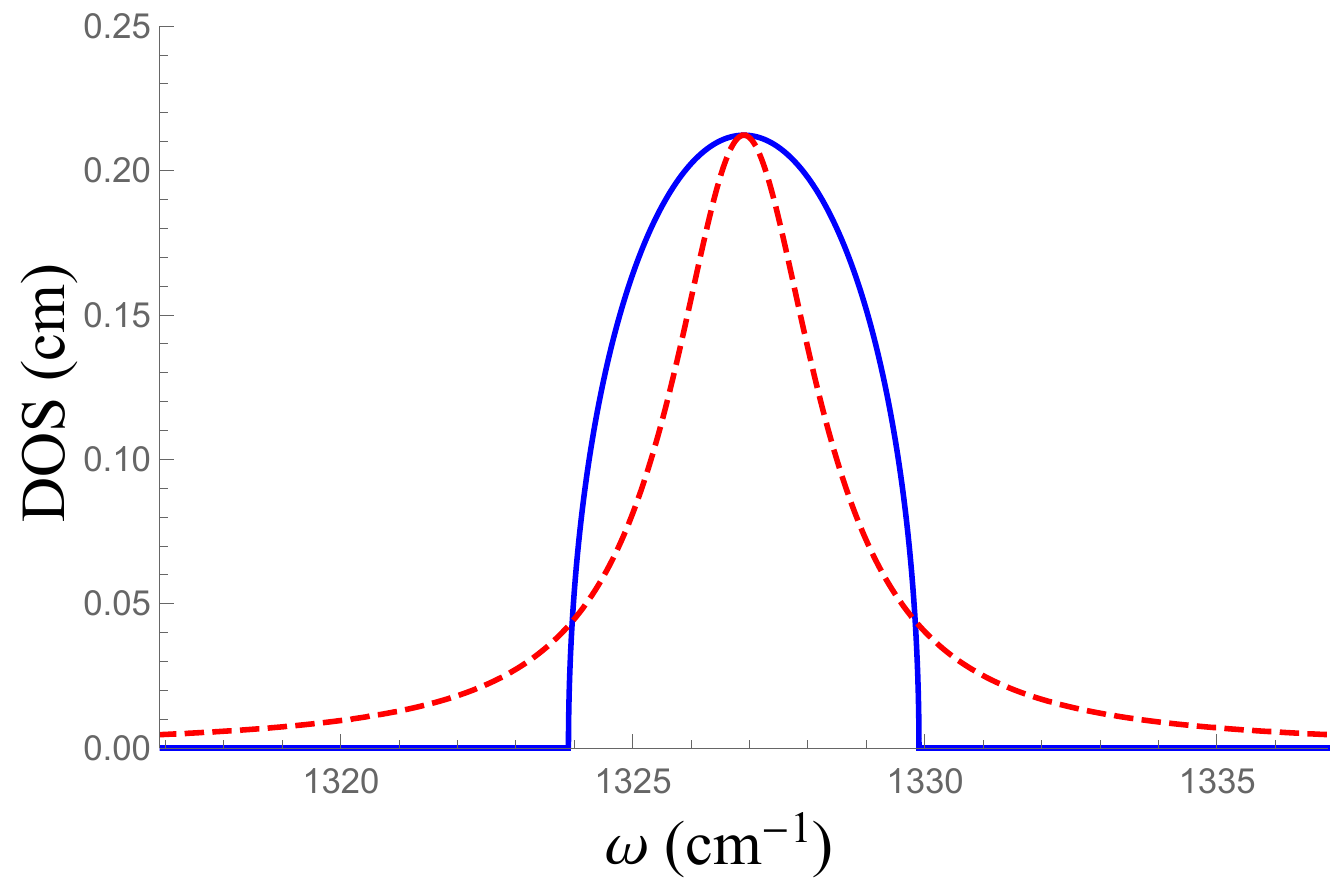}
  \caption{The shape of a spectral line in the regime of separated levels (blue line) vs. its on-shell counterpart  (red dashed line).
\label{SeparLine}}
\end{figure}

Semi-circle shape of a spectral line in the independent levels approximation and its simplified Lorentzian form are shown in Fig.~\ref{SeparLine}.

Next, we express the number of atoms $N$ in a particle via its effective size $L$. For example, cubic particles provide $N \rightarrow \, (\, 2L \, / \, P_6 \, a_0 \,)^3$, the lattice assumed to be cubic with two atoms in the unit cell. This finally yields:
\begin{equation}\label{GDMM2}
  \Gamma_n  =  \omega_n \, \mu_n(p) \sqrt{S} \left( \frac{a_0}{L} \right)^{3/2}.
\end{equation}
Here we absorbed the shape-dependent parameter $P_p$ into $\mu_n(p)= c_n (p) \sqrt{P^3_p/8}$. The constant $\mu_n(p)$ can be calculated analytically for cubic, spherical and cylindrical particles and numerically for any other shape of particles.

The real correction to the phonon self-energy stemming from other energy levels can be found from Eq.~\eqref{PiQW} without self-consistency:
\begin{equation}\label{PiQW3}
  \Pi_n (\omega) = \frac{S \omega_n}{8} \sum_{l, n^\prime \neq n}  \,\frac{  \left[ \mathbf{Y}_n(\mathbf{R}_l) \cdot \mathbf{Y}_{n^\prime}(\mathbf{R}_l) \right]^2 \,  \omega^2_{n^\prime}}{ \omega^2 - \omega^2_{n^\prime}} .
\end{equation}
This weakly $\omega$-dependent correction originates from the whole bunch of phonon modes in the spectrum and results mainly in the effective renormalization of $\omega_0$  for highest (i.e., closest to the Brillouin zone center) modes.

\subsection{Overlapped levels}
\label{SSWeakOver}

The difference between separated and overlapped levels is evident already when comparing Eq.~\eqref{PiQ} and Eq.~\eqref{Pi}. First, the sum over $n^{\prime}$ in Eq.~\eqref{PiQ} could be safely omitted within the independent levels approximation while the integration over intermediate momenta in Eq.~\eqref{Pi} is an essential ingredient of the theory. Second, the non-zero linewidth for separated levels appears only due to \emph{self-consistent} treatment while in the continuum the phonon line broadening is provided by the regular perturbation theory in $S$. In the latter case one gets:
\begin{equation}\label{ImPi}
{\rm Im} \, \Pi_{\bf q} (\omega) = -\frac{\pi S \omega_{\bf q}}{16 N} \sum_{{\bf k}} \omega_{\bf k} \left[ \delta (\omega + \omega_{\bf k)} + \delta (\omega - \omega_{\bf k}) \right].
\end{equation}
Eq.~\eqref{ImPi} contains two delta functions. Since we are interested in its behavior in the vicinity of positive
pole $\omega \approx \omega_{\bf q}$ of the Green's function, the phonon dispersion does not allow the
argument of delta-function $\delta (\omega + \omega_{\bf k})$ to reach zero value anywhere in the Brillouin
zone, so this term gives zero contribution to the integral. Hence, the only delta-function which contributes is
$\delta (\omega - \omega_{\bf k})$. Notice that in the vicinity of the
negative pole $\omega \approx - \omega_{\bf q}$ the situation is opposite. Then using the expansion of phonon spectrum near
its maximum one obtains:
\begin{equation}\label{ImPi1}
  {\rm Im} \, \Pi_{\bf q} (\omega) = -\frac{ S \, a_0^3 \, \omega_{\bf q}}{64 \, \pi \, \alpha^{3/2} } \,   \theta(\omega_0 - \omega)\, \omega \, \sqrt{\omega_0 - \omega}.
\end{equation}
We see that the imaginary part of $\Pi_{\bf q} (\omega)$ is nonzero only for frequencies lower than $\omega_0$. Moreover, it reveals the square-root
non-analytical behavior when $\omega \to \omega_0 $ originated from van Hove singularity in the phonon density of states
that occurs at this energy.

The real part of  $\Pi_{\bf q} (\omega)$ can be calculated on similar manner, and the result is given by
\begin{equation}\label{RePi}
{\rm Re} \, \Pi_{\bf q} (\omega) = -\frac{ S \, a_0^3 \, \omega_{\bf q}}{64 \, \pi \, \alpha^{3/2} } \, \theta(\omega - \omega_0) \, \omega \, \sqrt{\omega - \omega_0}.
\end{equation}
This result is essentially frequency-dependent in the vicinity of the phonon pole
$\omega \simeq \omega_{\bf q} \approx \omega_0 - \alpha q^2$. Likewise the contribution Eq.~\eqref{ImPi1} to the imaginary part of $\Pi_{\bf q} (\omega)$
it stems from  van Hove singularity in spectra of intermediate phonons scattered by disorder. It can be evaluated either directly or from the imaginary part of $\Pi_{\bf q} (\omega)$
via Kramers-Kronig relations, see~\cite{agd,landauV}.

In addition to Eq.~\eqref{RePi}, there appears one more term almost constant near the pole $\omega \approx \omega_{\bf q}$ and proportional to the Debye momentum $q_D$ that
comes from the upper limit of integration over the Brillouin zone in Eq.~\eqref{Pi}. It is the momentum domain where the details of phonon band structure become important, and the Debye approximation is valid only qualitatively. Moreover, the Keating model we use in our calculations essentially fails in this area. This makes the numerical prefactor in front of the high-momenta contribution very unreliable.   Also, similar contributions stems in this case from both $(\omega \pm \omega_{\bf k})^{-1}$ poles of the integrand in Eq.~\eqref{Pi}.
In fact, this term has the same origin as the contribution discussed below for certain types of disorder with nonzero mean value $ \langle \delta m_l^{-1} \rangle$ which provides the constant shift of the real part of phonon self-energy. Below we shall
concentrate on strongly frequency- and momentum- dependent part of the self-energy assuming that all constant (${\bf q}, \omega$-independent) terms are already absorbed into $\omega_0$ (see, however, Section~\ref{SStrong} and paper II).

\begin{figure}
  \centering
  \includegraphics[width=6.4cm]{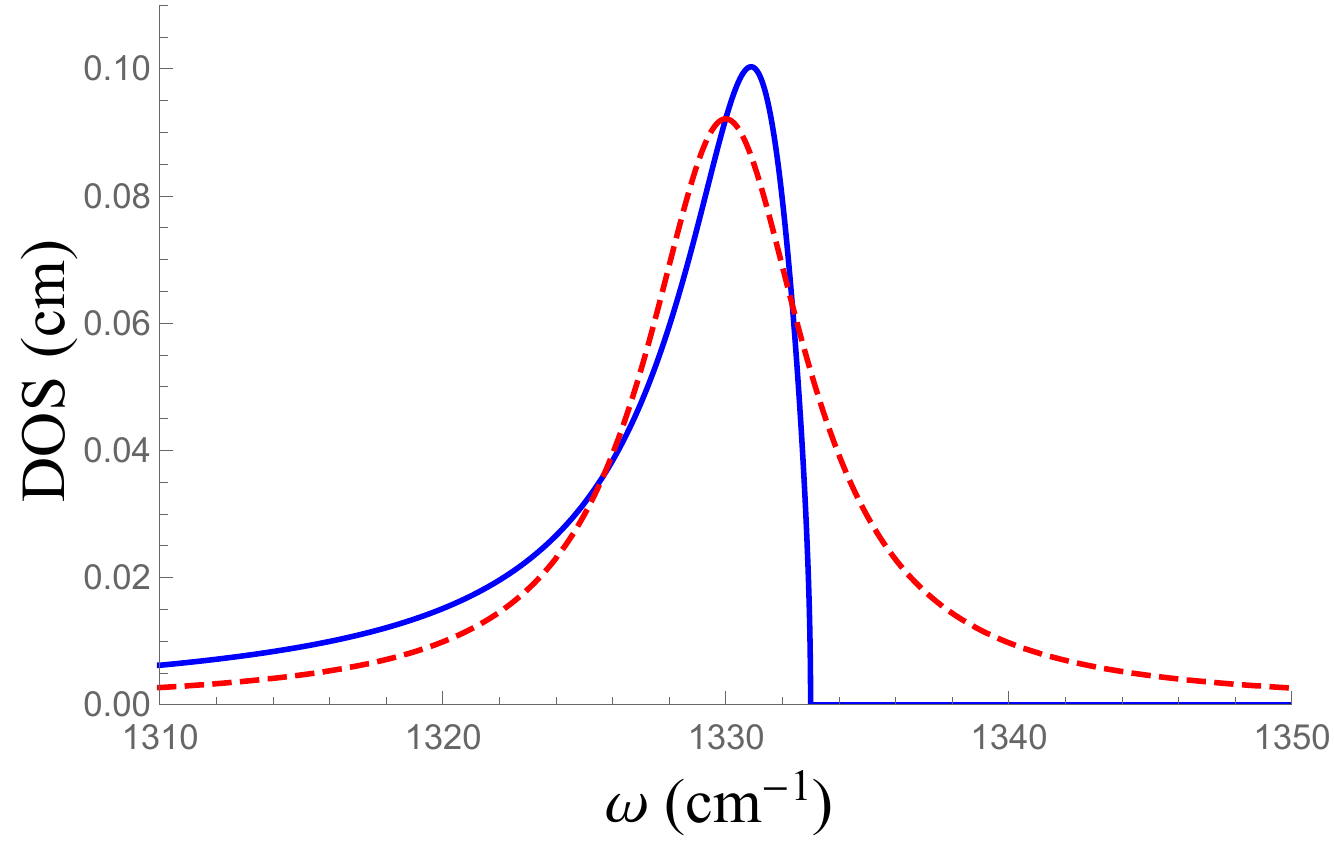}
  \caption{The shape of a spectral line in the regime  of overlapped levels (blue curve). The spectral weight is essentially asymmetric. Conventional Lorentzian shape (red dashed curve) can be obtained using the on-shell approximation.
\label{OverLine}}
\end{figure}

Notice that the real and the imaginary parts of
$\Pi_{{\bf q}} (\omega)$ can be conveniently rewritten jointly as follows:
\begin{equation}\label{Pi2}
\Pi_{\bf q} (\omega) =-\frac{ S \, a_0^3 \, \omega_{\bf q}}{64 \, \pi \, \alpha^{3/2} } \, \omega \, \sqrt{\omega - \omega_0}.
\end{equation}
Substituting Eq.~\eqref{Pi2} into Eq.~\eqref{D} we can draw
the spectral weight of the phonon Green's function
\begin{equation}\label{A}
A_{\bf q}(\omega)= \frac{1}{\pi} {\rm Im} D_{\bf q} (\omega).
\end{equation}
This function depicted in Fig.~\ref{OverLine} (blue curve) represents the broadened spectral line of the optical phonon (rigorously speaking, in
the bulk). Our first observation is that the shape of the line is quite asymmetric. This asymmetry originates from nontrivial frequency dependencies of both real and imaginary parts of
the self-energy which take place in different frequency domains. It is the common feature in many physical
problems where the self-energy demonstrates visible frequency variation near the quasiparticle pole (see, e.g., Ref.~\cite{toperverg1993}). Let us perform, however, the {\it on-shell approximation}:
\begin{eqnarray}\label{Loren}
\Gamma_{\bf q}(\omega) = - \frac{2 \omega_{\bf q} {\rm Im} \Pi_{\bf q} (\omega)}{\omega + \omega_{\bf q}} \, \rightarrow \, \Gamma_{\bf q}(\omega = \omega_{\bf q})   \equiv  \Gamma_{\bf q}.
\end{eqnarray}
Making this substitution in Eq.~\eqref{A} we observe that the resulting
curve becomes Lorentzian (see Fig.~\ref{OverLine}, red dashed line), i.e., symmetric near its maximum, the width of the peak being close to the real one {\it
as far as the interaction-induced effects are small}.

Ignoring the lineshape asymmetry yields the phonon linewidth in the form
\begin{equation}\label{G1}
\Gamma_{\bf q} = \omega^2_{\bf  q} \,\, \frac{S \, a_0^3}{64 \pi \alpha } \,  q = \omega_{\bf q} \,\,
\frac{S }{64 \pi F} \, ( q \, a_0 ).
\end{equation}
The linear-in-$q$ dependence of the broadening parameter $\Gamma_{\bf q}$ here is very important. Until now our consideration was related to the bulk
phonon-impurity model. Now we perform the finite-size quantization $q \rightarrow q_n(L, p)$ in Eq.~\eqref{G1}.
It immediately yields the linewidth dependence on the particle size in the desirable~\cite{yoshikawa1993raman}
$1/L$ form:
\begin{equation}\label{G2}
\Gamma_n = \omega_{n} \, \nu_n(p) \, S \, \frac{a_0}{L}.
\end{equation}
Here $\nu_n (p)$ is the shape-dependent coefficient which contains: (i) numerical prefactor $1/64$, (ii) spectrum flatness parameter $F$ in denominator, and (iii) strong dependence on the quantum number $n$. From the above consideration it follows that the phonon modes with large $n$ are broadened stronger than the first one. In particular, for cubic particles their widths are related to each other as
\begin{equation}\label{GG3}
\Gamma_{n_{x},n_{y},n_{z}}= \sqrt{\frac{n_x^2 + n_y^2 + n_z^2}{3}} \, \Gamma_1.
\end{equation}
Notice that for Raman active modes all quantum numbers $n_i$ must be odd in this case.

To conclude this Section, we investigated how the simplest disorder (weak point-like delta-correlated impurities) affects vibrational modes
in nanoparticles. We calculated the phonon linewidths in regimes of separated and overlapped levels. For separated levels we used the self-consistent Born approximation and observed the semi-circle law for the phonon density of states and $\sqrt{S}/L^{3/2}$ impurity strength and particle size dependence of the linewidth, the prefactor is found to be weakly shape and quantum number dependent quantity. For overlapped levels we found that the phonon line shape is asymmetric due to van Hove singularity in the spectrum. By applying the Lorentzian approximation
and further quantizing the phonon momentum we obtain the damping proportional to $S/L$ with prefactor strongly dependent on the phonon quantum number.


\section{Smooth random potential}
\label{SSmooth}

Section~\ref{SWeak} has been devoted to the treatment of weak point-like (delta-correlated) Gaussian disorder. Now we shall consider how the smooth character of the (still, weak) random impurity potential affects the results of previous Section. More specifically, instead of point-like impurities Eq.~\eqref{delta m 2 av} we shall assume the long-range character of the impurity-impurity correlation function
\begin{multline}\label{delta m 2 av smooth}
\langle \,\delta m^{-1}_l \, \delta m^{-1}_{l^{\prime}} \, \rangle \,\, m^2 \, = \\
\langle \,\delta m_l\, \delta m_{l^{\prime}} \, \rangle \,\, m^{-2} \, = \\
S \,  W \left(|\mathbf{R}_l - \mathbf{R}_{l^{\prime}}|; \sigma \right),
\end{multline}
where the characteristic scale $\sigma$ of the potential is much longer than the lattice parameter, $\sigma \gg a_0$, and the correlation function is Gaussian:
\begin{equation}\label{smooth}
W (r; \sigma)= \frac{a_0^3}{(2 \pi \sigma^2)^{3/2}} \, \exp{\left(- \frac{r^2}{2 \sigma^2} \right) },
\end{equation}
the prefactor is introduced for proper normalization.

Notice that particular spatial dependence of correlator in Eq.~\eqref{delta m 2 av smooth} is not essential; the only
property needed is its rapid decay on scale $\sigma$. For instance, we observe that replacing Gaussian correlator in Eq.~\eqref{smooth} by the exponential one does not change qualitatively our results.

\subsection{Separated levels}
\label{SSSmoothSep}

Here we present our results for separated levels. We obtain the self-consistent equation for the phonon self-energy in the following form:
\begin{eqnarray}\nonumber
  \Pi_n (\omega)  &=&  \frac{S \, \omega_n}{8} \! \sum_{l,l^\prime, n^\prime}  \!  ( \mathbf{Y}_{n l} \cdot \mathbf{Y}_{n^\prime l}) \, W_{l  l^{\prime}} (\sigma) \, (\mathbf{Y}_{n l^\prime} \cdot \mathbf{Y}_{n^\prime l^\prime}) \\ \label{PiQS}
  && \,\, \quad\qquad\qquad  \times \, \frac{\omega^2_{n^\prime}}{ \omega^2 - \omega^2_{n^\prime} - 2 \omega_{n^\prime} \Pi_{n^\prime}(\omega)},
\end{eqnarray}
where we used the shorthand notations $\mathbf{Y}_{nl} = \mathbf{Y}_{n} (\mathbf{R}_l)$ and $W_{l  l^{\prime}} (\sigma) =  W \left(|\mathbf{R}_l -\mathbf{R}_{l^{\prime}}|; \sigma \right)$. For non-degenerate phonon levels it yields:
\begin{equation}\label{PiQS1}
  \Pi_n (\omega) = \frac{2 \, d^{\, 2}_{n} (p) \, S \, \omega^3_n}{N}  \,\frac{ 1}{ \omega^2 - \omega^2_{n} - 2 \omega_{n} \Pi_{n}(\omega)},
\end{equation}
where parameters $d_n (p)$ are given by
\begin{equation} \label{CS1}
  d^{\, 2}_{n} (p) = \frac{N}{16} \sum_{l,l^\prime}  \mathbf{Y}^{\, 2}_{n l} \, W_{l l^\prime} (\sigma) \, \mathbf{Y}^{\, 2}_{n l^\prime}.
\end{equation}
These constants could be calculated for any particular model and particle shape. Up to new prefactors $d_n(p)$ instead of $c_n(p)$ all the results of Subsection~\ref{SSWeakSep} are applicable for the case of a smooth random potential. Once again, using the on-shell approximation we obtain the line broadening in the form:
\begin{equation}\label{GS2}
  \Gamma_n  = \omega_n \, \rho_n(p) \,\sqrt{S} \, \left( \frac{a_0}{L} \right)^{3/2},
\end{equation}
where $\rho_n(p)$ are the novel shape and quantum number dependent parameters, $\rho_n(p)= d_n (p) \sqrt{P^3_p/8}$.

The smooth character of disorder leads to diminishing of parameter $\rho_n(p)$ in comparison with $\mu_n(p)$ (see Appendix). Two regimes should be distinguished. The first one could be analyzed via an expansion in small parameter $q_n  \sigma \ll 1$, where $q_n (L,p)$ is the quantized discrete momentum, $\omega_n =\omega_0 - \alpha  q^{\, 2}_{n}$. The leading order contribution coincides with the result for point-like impurities, the first correction being of the order of $(q_n \sigma)^2$.

The second regime that occurs at $q_n \sigma \gg 1$ appears exclusively due to smoothness of the random potential. In this regime the wave-functions in Eq.~\eqref{CS1} experience fast oscillations on the scale $\sigma$ and therefore could be replaced by their average values $\sim 1/\sqrt{N}$. As a result, the linewidth as a function of particle size $L$ saturates:
\begin{equation}\label{GS3}
   \Gamma_n = \omega_n \, \frac{1}{4(2\pi)^{3/4}} \, \sqrt{S} \, \left( \frac{a_0}{\sigma} \right)^{3/2}.
\end{equation}
The damping in this regime is small as compared to the delta-correlated case (cf. Subsection~\ref{SSWeakSep}).

\subsection{Overlapped levels}
\label{SSSmoothOver}

The leading contribution to ${\rm Im} \, \Pi_{\bf k} (\omega)$ is determined by the same diagram Fig.~\ref{Diagrams} as for delta-correlated impurities.
The corresponding analytical expression reads:
\begin{equation}\label{ImPi4}
{\rm Im} \, \Pi_{\bf q} (\omega) = -\frac{ \pi S \omega_{\bf q} }{16 N} \sum_{{\bf k}} \, \omega_{\bf k} \widetilde{W}({\bf k - q}; \sigma)  \delta (\omega - \omega_{\bf k}),
\end{equation}
where $\widetilde{W}({\bf q}; \sigma)$ is the Fourier transform of Eq.~\eqref{smooth}:
\begin{equation}\label{smooth-K}
\widetilde{W}({\bf q}; \sigma)= \exp \left( - \frac{q^2 \sigma^2}{2} \right).
\end{equation}
Integrating in Eq.~\eqref{ImPi4} one obtains:
\begin{eqnarray}\label{GammaS1O} \nonumber
\Gamma_{\bf q} (\omega) &=&  -\frac{ S \, a_0^3 \, \omega_{\bf q} \, \omega }{64 \, \pi }  \, \frac{\theta (\omega_0 - \omega)}{2 \, \alpha \, \sigma^2 \, q} \\ \nonumber
&\times&
\left( \exp{ \left[- \frac{1}{2} \frac{\sigma^2}{\alpha} ( \sqrt{\alpha} q - \sqrt{\omega_0 - \omega} )^2 \right] } \right. \\
&-&
\left. \exp{ \left[- \frac{1}{2} \frac{\sigma^2}{\alpha} ( \sqrt{\alpha} q + \sqrt{\omega_0 - \omega} )^2 \right] } \right).
\end{eqnarray}
The next step is to apply
for Eq.~\eqref{GammaS1O} the on-shell approximation $\omega = \omega_{\bf q}$. It yields:
\begin{equation}\label{GammaS2O}
\Gamma_{\bf q} \, = \, \Gamma^{0}_{\bf q} \, {\rm f} \, ( q \sigma ),
\end{equation}
where $\Gamma^{0}_{\bf q} $ is the bulk phonon damping for delta-correlated disorder
(see Eq.~(\ref{G1})), and the spreading function ${\rm f} \, (x) $ is given by the following expression:
\begin{equation}\label{f}
{\rm f} \, (x) = \,    \frac{1-\exp{[- 2 x^2]} }{2 x^2}
\end{equation}
Now we ready to apply the finite size quantization replacement $q \to q_n (L , p)$ in Eqs.~(\ref{GammaS2O}) and (\ref{f}):
\begin{equation}\label{GammaSO}
\Gamma_n  \, = \, \Gamma^{0}_n  \, {\rm f} \, ( q_n \sigma ),
\end{equation}
where $\Gamma^{0}_n $ is given by Eq.~(\ref{G2}). This equation determines the damping of overlapped vibrational eigenmodes in finite nanoparticles subjected to the smooth random potential for arbitrary relation between $L$ and $\sigma$.

It is instructive to investigate Eq.~\eqref{GammaSO} in limiting cases of small $q_n \sigma \ll 1$ and large $q_n \sigma \gg 1$ discrete momenta.
The first asymptote yields ${\rm f} \, (x) \to 1$ and therefore $\Gamma \approx \Gamma^{0}$; in the leading order scale $\sigma$ drops down from the formula. Thus, the damping function for smooth disorder qualitatively reproduces its behavior peculiar for point-like impurities. We conclude that for finite nanoparticles which are essentially inhomogeneous on the scale of its size, $L \geq \sigma$, the model of point-like impurities qualitatively describes the case of smooth disorder, as well, and the resulting linewidth  still mainly follows $\Gamma \propto 1/L$ law.

\begin{figure}
  \centering
  \includegraphics[width=6.4cm]{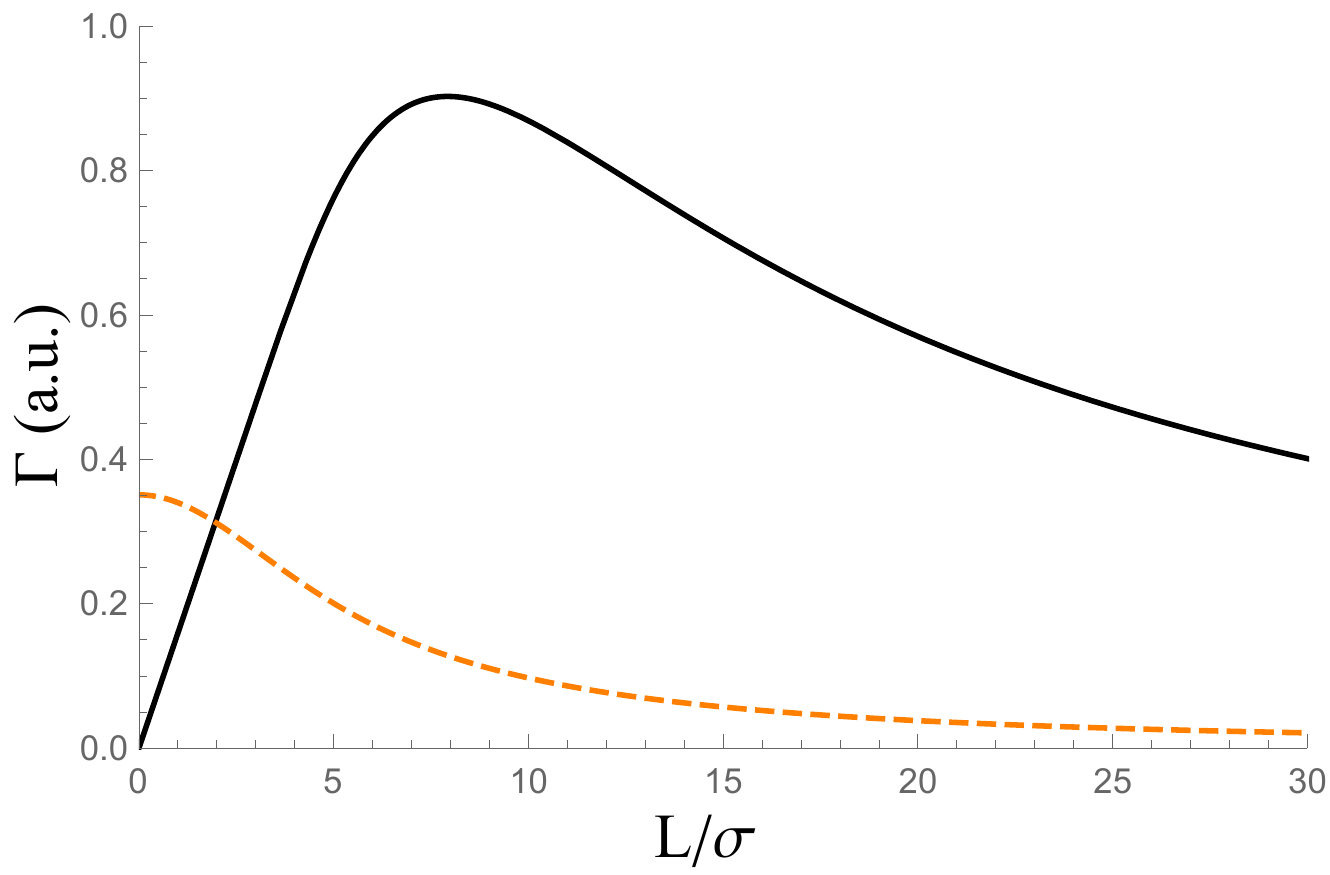}
  \caption{Sketch of the phonon linewidth dependence on the particle size $L$, with $\sigma$ being the characteristic length scale of the smooth random potential. Black line depicts the damping of the first phonon eigenmode ($q_1 \approx 2 \pi/L$ in Eq.~\eqref{GammaSO}) in the regime of overlapped levels. At small $L$ the damping behaves as $\Gamma \propto L$, while for large $L$ one observes the dependence peculiar for point-like impurities, $\Gamma \propto 1/L$. In the regime of separated levels (orange dashed curve) the damping monotonically decreases from finite value at small $L$ towards zero as $\Gamma \propto 1/L^{3/2}$ when $L$ increases.
\label{Smooth}}
\end{figure}

However,  when the discrete momentum is large,  ${q_n \sigma \gtrsim 1}$, then ${\rm f} \, (x) \approx 1/(2 x^2)$, and the asymptotic behavior of damping in smooth disorder described by Eq.~\eqref{GammaSO} becomes $\Gamma_n \propto 1/q_n \sigma^2$ thus being drastically different from the Born case. In terms of
finite-size particles it means that the long-range character of spacial variations of disorder manifests itself only when it occurs
on scales of order of or longer than the particle size, $\sigma \geq L$, when the novel regime $\Gamma \propto L$ emerges:
\begin{equation}\label{GammaSO1}
\Gamma_n  \, = \, \frac{\omega_n}{2} \, \left(\frac{1}{64 \pi F} \right)^2 \, \left(\frac{a_0}{\sigma}  \right)^2 \, \nu^{-1}_n (p) \, S  \, \frac{L}{a_0} .
\end{equation}
Physically, two regimes of momentum dependence for $\Gamma_{\bf q}$ could be understood from the general scattering theory~\cite{landauIII} after we recognize that the particle velocity ${\bf v}_{\bf q} \propto \nabla_{\bf q} \, \omega_{\bf q}$ is proportional to its momentum $q$. Then the first regime corresponds to the scattering of ``slow'' particles and the second one describes the ``fast'' particles scattered by soft-core spheres with radii $\sigma$ (see the problems in Ref.~\cite{landauIII}).

This Section has been devoted to the treatment of effects of smooth character of the impurity potential. Introducing the characteristic potential
scale $\sigma$ we observed, that the relation between the nanoparticle size $L$ and the scale $\sigma$ is responsible for two crossovers in the behavior of the phonon linewidth.
Namely, when the particle size is larger than the impurity length scale, i.e., when the nanoparticle is essentially inhomogeneous, the damping reveals ordinary $1/L^{3/2}$ dependence for separated and  $1/L$ dependence for overlapped phonon levels. Otherwise, the particle could be regarded as ``almost homogeneous''. Then the damping  behavior changes to saturation for separated and to linear-in-$L$ dependence for overlapped levels, both crossovers occur when $L \sim \sigma$ (to be precise, the numerical factor $2 \pi$ entering expansions moves  crossovers towards larger $L/\sigma$ values, see Fig.~\ref{Smooth}).

\section{Strong impurities}
\label{SStrong}

In this Section we investigate the influence of strong rare impurities on the spectrum of optical vibrations in particles. We utilize the standard T-matrix approach (see, e.g, Ref.~\cite{doniach1998}) which treats exactly multiple scattering off a single impurity, see Fig.~\ref{DiagScat}c. It  gives the correct result to the first order in $c_{imp} \ll 1$ . The possibility to form the phonon-impurity bound state is also analyzed.

More specifically, instead of weak (Born Gaussian distributed) impurities with $ (\delta m/m)^2 \ll 1$ and zero average $\langle \delta m_l \rangle = 0$ we introduce the binary distributed strong disorder (arbitrary $\delta m \geq -m$). The mass variation reads:
\begin{equation}\label{Tpert}
  \delta m_l = \sum_{l^\prime} \delta_{l l^\prime} \, \delta m ,
\end{equation}
where the sum runs over all impurity atoms located at sites $l^\prime$; the masses of host atoms and impurity atoms are given by $m$ and $m + \delta m$, respectively.

We also introduce the dimensionless impurity potential
\begin{equation}\label{TImp}
  U = \frac{\delta m}{m + \delta m},
\end{equation}
which will be useful for calculations below.

\subsection{Separated levels}
\label{SSStrongSep}

The T-matrix approximation presumes exact solution of the single-impurity problem with its subsequent averaging over disorder configurations, the problem should be solved self-consistently. Let us discuss the first step. Contribution to $\Pi_n (\omega)$ stemming from all multiple scattering off the impurity located at the $l$-th site reads:
\begin{eqnarray}\label{SigmaT1}
  \Pi_n^{\, l} (\omega) &=& \omega_n \sum^\infty_{k=1} \sum_{n_1, \ldots , n_{k-1}} \left(-\frac{U}{4}\right)^k ( \mathbf{Y}_{n l} \cdot \mathbf{Y}_{n_1 l} )  \nonumber \\
  &\times& ( \mathbf{Y}_{n_1 l} \cdot \mathbf{Y}_{n_2 l} ) \ldots ( \mathbf{Y}_{n_{k-1} l} \cdot \mathbf{Y}_{n l} ) \nonumber \\
  &\times & \omega_{n_1} D_{n_1}(\omega) \ldots \omega_{n_{k-1}} D_{n_{k-1}}(\omega),
\end{eqnarray}
the $k$-th term corresponds to the $k$-fold scattering off this impurity. As the next step, one should sum up the r.h.s. of Eq.~\eqref{SigmaT1} over all impurity positions.

The independent levels approximation allows to solve this problem by putting all $n_i$ in Eq.~\eqref{SigmaT1} equal to $n$. Simple argument $|\mathbf{Y}(\mathbf{R})| \sim 1/\sqrt{N}$ allows to estimate the $k$-th contribution to Eq.~\eqref{SigmaT1} as  $1/N^k$. Thus, at $ |U| \sim 1$ it is sufficient to restrict our consideration by two first terms in Eq.~\eqref{SigmaT1} corresponding to diagrams (a) and (b) in Fig.~\ref{DiagScat}.  The first one is the (size-independent) rescaling of the phonon frequency $\omega_n$ due to disorder-induced change of the average mass of atoms (cf. paper II):
\begin{equation}\label{SigmaT2}
  \Pi^{(1)}_n = - \frac{1}{4} \, \omega_n \, U \, c_{imp}.
\end{equation}

The second term in Eq.~\eqref{SigmaT1} results in the self-consistent equation for $\Pi_n(\omega)$ which is equivalent to Eq.~\eqref{PiQW1} where $S$ one should by replaced by $c_{imp} \, U^2$. The counterpart of Eq.~\eqref{GDMM2} reads:
\begin{equation}\label{GDMMT}
  \Gamma_n  =  \omega_0 \, \mu_n(p) \, \sqrt{c_{imp}} \,  \left| \, U  \right| \, \left( \frac{a_0}{L} \right)^{3/2}.
\end{equation}

Addressing the problem of localization of optical phonons by impurities we notice that for $\delta m < 0 $ this localization can appear at high frequencies $\omega > \omega_0$ because lighter atoms have higher vibrational eigenfrequencies. The corresponding analysis taking into account all phonon modes $n_i$ in Eq.~\eqref{SigmaT1} and summing up all contributions of different orders in $k$ can be realized for any particular set of model wave functions $\mathbf{Y}_n$ numerically but hardly representable in general terms analytically. Moreover, the capability of impurity to localize  phonons turns out to be strongly dependent on its location in a particle (see paper II). Therefore, we postpone the detailed discussion of localization issues to the next Subsection where the boundary induced phenomena are excluded from consideration due to the ``bulk'' approach we use and to the paper II where they addressed numerically in the most general form.

\subsection{Overlapped levels}
\label{SSStrongOver}


In the bulk all contributions shown in Fig.~\ref{DiagScat}c can be easily summed up as a geometrical progression. After averaging over disorder the phonon self-energy acquires the following standard form:
\begin{equation}\label{PiSt1}
\Pi_{\bf q} (\omega) = - \frac{1}{4} \,\omega_{\bf q} \, \frac{\displaystyle{c_{imp} \,  U}}{\displaystyle{1 + \frac{U}{4 N } \, \sum_{\bf k} \, \omega_{\bf k} \,  D_{0} (\omega,  {\bf k})}}.
\end{equation}
It is worth to mention that $\Pi_{\bf q} (\omega)$ in Eq.~(\ref{PiSt1}) remains finite even for infinitely strong $U$ (``unitary limit'').

Integration in Eq.~\eqref{PiSt1} is performed similarly to the Born case. One important difference comes from the fact that due to different model of disorder
(binary disorder with nonzero average) now we can not absorb constant terms into $\omega_0$. As a result we get:
\begin{equation}\label{Den}
\frac{1}{N} \sum_{\bf k} \, \omega_{\bf k} \, D_{0} (\omega, {\bf k}) = \frac{\omega_0}{4 \pi \alpha} \left[ \frac{q_D}{\pi /2} - \sqrt{\frac{\omega - \omega_0}{\alpha}} \right],
\end{equation}
where $q_D$ is the Debye momentum (upper limit of integration). The quantity $q_D$ is in fact a sort of adjustable parameter since its definition presumes that the spherical symmetry of the long wavelength phonon spectrum could be extended onto large momenta. Therefore, the factor $\pi/2$ in Eq.~\eqref{Den} could not be treated seriously, and we just absorb it into the definition of Debye momentum using $q_{D}^{\prime} = q_D / (\pi /2 )$ instead of $q_D$.

Plugging Eq.~\eqref{Den} into Eq.~\eqref{PiSt1}, at $\omega<\omega_0$ we obtain:
\begin{equation}\label{PiStrong1}
\Pi_{\bf q}(\omega)    = - \, \omega_{\bf q}  \, 4  \pi F^2 \,  \frac{ c_{imp} \, \omega_0}{\omega_{loc}-\omega} \, \left[\, \frac{a_0}{\zeta} \, +  \, i \, \sqrt{\frac{\omega_0 - \omega}{F \omega_0}} \, \right],
\end{equation}
where the resonant frequency $\omega_{loc}$ (see below) is determined as follows:
\begin{equation}\label{Res}
\omega_{loc} = \omega_0 + \alpha \zeta^{-2},
\end{equation}
and the characteristic length $\zeta$ is given by
\begin{equation}\label{zeta}
\zeta^{-1}= q_D^{\prime} \, \left( 1+ \frac{16 \pi F}{U q_D^{\prime} a_0}  \right).
\end{equation}
This length is of order of $1/q_D^{\prime}\sim a_0$ for regular $U$ and infinitely increases when $U \to U_{min} = - 16 \pi F / q_D^{\prime} a_0 $.
At $\omega > \omega_0$ we get ${\rm Im} \, \Pi_{\bf q}(\omega)=0$, i.e., the damping is absent.

Consider the on-shell  approximation $\omega = \omega_{\bf q}$. The damping acquires the form:
\begin{equation}\label{GammaStrong1}
\Gamma_\mathbf{q} =  \, \omega_{\bf q} \, 4 \pi F \, c_{imp} \, \left(\frac{\zeta}{a_0} \right) \, \frac{q \zeta}{1 + (q \zeta)^2}.
\end{equation}
After applying the finite size quantization replacement $q \to q_n \, (L,p)$ it yields:
\begin{equation}\label{GammaStrong2}
\Gamma_n  =  \, \omega_n \, 4 \pi F \, c_{imp} \, \left(\frac{\zeta}{a_0} \right) \, \frac{q_n \zeta}{1 + (q_n \zeta)^2}.
\end{equation}
We see that the behavior of the linewidth depends on parameter $\zeta$. When the spatial scale generated by this quantity is short, $\zeta \sim a_0$, the linewidth reveals conventional $1/L$ behavior. If, however, $\zeta$ is a sort of critical quantity (this happens when $U \approx U_{min}$), than it generates the long scale to be compared with $L$.  For large  $L \gg \zeta$ Eq.~(\ref{GammaStrong2}) yields $1/L$ particle size dependence with additional enhancement factor $(\zeta / a_0 )^2$. In the opposite case $L \ll \zeta$ the size dependence in Eq.~(\ref{GammaStrong2}) changes to linear-in-$L$, the situation resembles crossover observed for smooth random potential in Section~\ref{SSmooth}.

Let us present the formula for the phonon linewidth in the unitary limit $U \to -\infty $ (or $\delta m \approx -m $, very light atoms or free vacancies):
\begin{equation}\label{GBulkT2}
  \Gamma_n  = \omega_n \, \left( \frac{32 F}{a_0  q_D} \right)^2 \, c_{imp}      \, \nu_n (p) \, \frac{a_0}{L},
\end{equation}
the strength of disorder does not enter this result.
In contrast, for large positive $\delta m \gg m$ (very heavy impurities) disorder potential saturates at $U \to 1$ and there is no unitary limit.

\begin{figure}
  \centering
  \includegraphics[width=6.4cm]{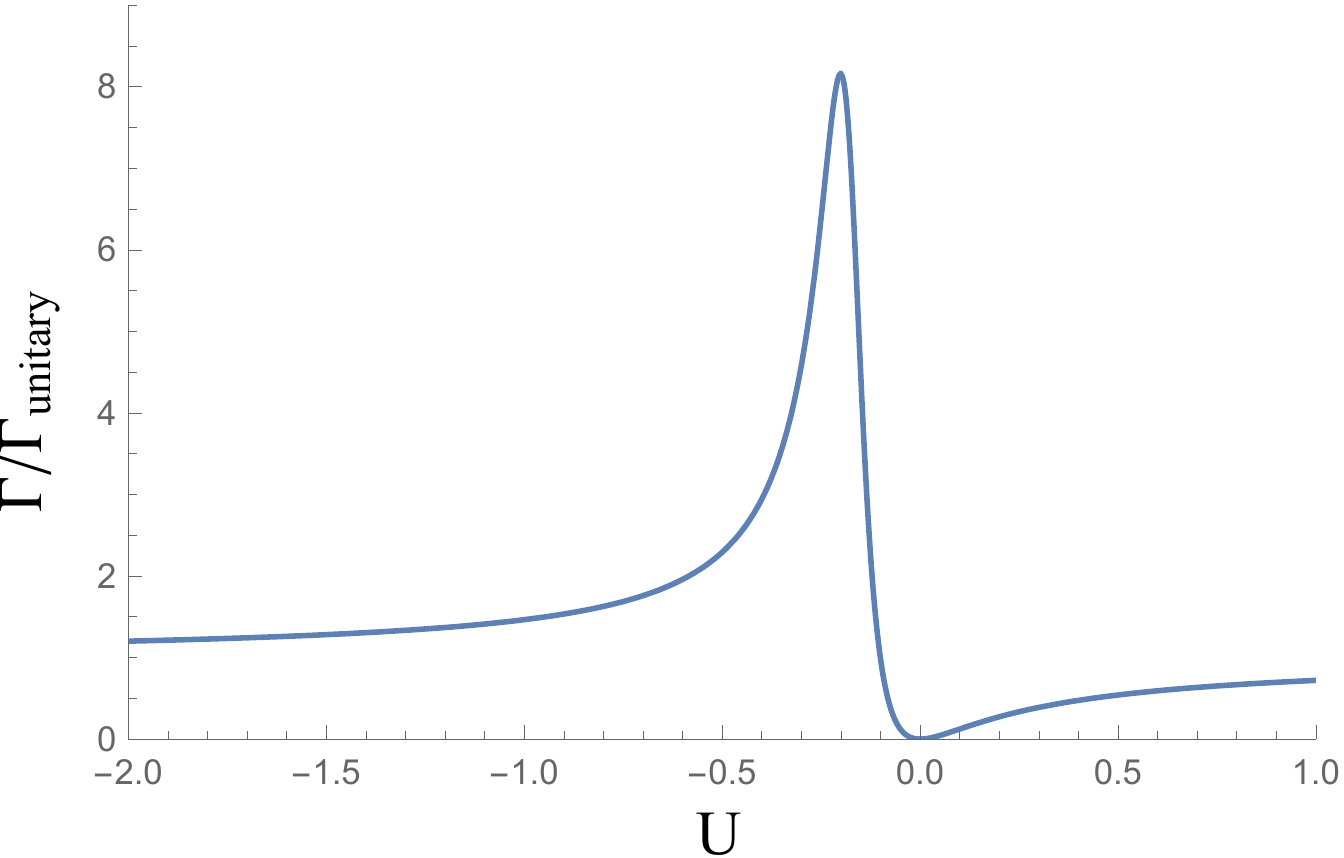}
  \caption{Plot of the first phonon mode linewidth dependence on the disorder potential $U$ in the regime of overlapped levels for a spherical 3nm nanodiamond.
   We normalize the damping to its unitary limit which corresponds to $U \to -\infty $ or $\delta m \approx -m$ (very light impurities or vacancies). At small $U$ the line width is proportional to $U^2$ as in the case of weak point-like impurities. At negative $U$ near the threshold of appearance of isolated level $U_{min} \approx -0.2$ the damping dramatically increases due to resonant scattering. For phonons with higher momenta the resonance is much smoother.
\label{FigT}}
\end{figure}

In Fig.~\ref{FigT} we plot the linewidth dependence on the potential $U$ for the first phonon mode in spherical 3nm nanodiamonds. One can see the quadratic $U$-dependence at $|U| \ll 1$ which corresponds to weak point-like impurities (see Subsection~\ref{SSWeakOver}), the resonant scattering when $U$ is close to $U_{min} \approx -0.2$ and the unitary limit at $U \lesssim -1$, when the damping is almost constant.

Importantly, Eq.~\eqref{Den} allows to study the possibility to form phonon-impurity bound states by investigating zeros of denominator in Eq.~\eqref{PiSt1} which determines the resonant energies of these states. At $\delta m>0$ localized states with $\omega>\omega_0$ do not appear because the denominator is positive in this frequency domain. Significant perturbations in density of states may only be observed at frequencies corresponding to short-lived short-wavelength optical phonons or even below the optical band. These states are useless for the analysis of Raman spectra.

In the meantime, for light impurities $\delta m<0$ the situation is very different. At $U \leq U_{min} $ equation
\begin{equation}\label{TLoc1}
  1 + \frac{U a_0}{16 \pi F} \left( q^\prime_D - \sqrt{\frac{\omega -\omega_0}{\alpha}} \right)=0
\end{equation}
has a solution given by Eq.~(\ref{Res}). When $U \approx U_{min}$ the frequency of localized level $\omega_{loc}$ is close to $\omega_0$ and the phonon damping is drastically enhanced due to resonant character of scattering off the impurities. With further decreasing of $U$ towards the unitarity the main contribution to the integral in Eq.~\eqref{PiSt1} comes from momenta lying beyond the range of applicability of spectral expansion~Eq.~\eqref{omega 0} and Eq.~\eqref{Den} fails. Here we just mention that in the improved theory $\omega_{loc} \propto |U|^{1/2}$.

To conclude this Section, we investigate the optical phonon damping appearing in nanoparticles due to scattering off strong dilute point-like impurities within the framework of $T$-matrix approximation. In the regime of separated levels besides the trivial rescaling of the phonon eigenfrequency $\omega_n$ we observe the damping similar to the weakly disordered case: it is proportional to the square root of impurity concentration and decays as $L^{-3/2}$ with increasing of the particle size. For overlapped levels we find the linear-in-$q$ momentum dependence of the phonon damping in the bulk and convert it into $1/L$ dependence for finite-size particles. In the ``critical'' regime of resonant scattering we obtain the parametrical enhancement of $1/L$ damping for largest particles and crossover to linear-in-$L$ behavior for smallest ones. We examine in details the possibility to form the phonon-impurity bound state near the phonon band edge for both positive and negative impurity mass defects.


\section{Discussion and Intermediate Conclusions}
\label{SDisc}


This Section is intended to present the discussion of our analytical results elucidating the role of disorder in crystalline nanoparticles. More complex and detailed summary which includes the comparative analysis of both analytical and numerical approaches to this problem could be found in paper II. In the meanwhile, two general remarks are in order right now.

First, we would like to point out that in the above consideration we did not present the unified analytical theory of phonons in disordered nanoparticles. Instead, we investigate important features of this problem by studying several simplest models which probably exaggerate and oversimplify the real experimental situation (cf. paper II). It allows to extract characteristic spatial scales associated with these features and, more importantly, to formulate the {\it qualitative} model free picture of phonon-impurity scattering in nanoparticles.

Our second remark is related to the object we addressed in this paper. We would like to emphasize that we evaluated the broadening of {\it individual lines} in  spectra of vibrational eigenmodes in nanoparticles of a given shape. Only all these lines treated together, properly broadened, and elaborated with the use of the theory of photon-phonon interaction constitute the Raman peak in real experiments. This approach should be contrasted with PCM-like
theories wherein the linewidth is a single parameter for the entire Raman peak.

We started our analysis from the treatment of simplest weak point-like impurities. We observe that there exist two drastically different regimes depending on either the vibrational eigenmodes are separated or they belong to the (quasi)continuum. In particular, for separated levels the phonon damping behaves as $\Gamma_n \propto \sqrt{S}/L^{3/2}$, where $L$ is the particle size and the disorder strength $S$ is a product of impurity concentration $c_{imp}$ and the squared relative variation of the random parameter of the theory (in our case, the atomic mass $m$). For overlapped levels we find $\Gamma_n \propto S/L$. In the latter case we solve the
bulk problem and perform the finite-size quantization only in final formulas. This (definitely, approximate) approach is justified by evident similarity between the continuous spectra of propagating phonon modes and the ones of plane waves, the difference supposed to be negligible.

These results were obtained within the on-shell (Lorentzian) approximation. Beyond this approximation the line shape in overlapped regime is shown to be strongly asymmetric due to Hove singularity in the spectrum of optical phonons, whereas in separated regime it obeys the semi-circle law. We also establish simple relations between the linewidths of eigenmodes with different quantum numbers. The dependence of $\Gamma_n$ on the quantum number of the level is observed to be stronger in the overlapped regime than in the separated one.

Considering weak and smooth disorder with characteristic length scale $\sigma$  we obtain the damping at $\sigma \ll L$ similar to the one for point-like impurities. In the opposite limit $\sigma \gg L$ it is proportional to $L$ in overlapped regime and saturates as a function of particle size if the levels are separated. Such a behavior of the phonon linewidth for overlapped levels could be understood from general quantum mechanical picture of fast vs. slow particles scattered off the ``soft spheres'' potential~\cite{landauIII}.

At last, we treated the case of strong impurities with low concentration $c_{imp} \ll 1$ using the $T$-matrix approach borrowed from the theory of dirty fermions~\cite{doniach1998}. We consider the replacement impurities of a single sort to be strong so the effective local potential generated by mass defect may even exceed the phonon band width. Our main observation concerning separated levels is that the result for damping is almost the same as for the weak impurities case. The only difference is additional shift of the maximal optical phonon frequency $\omega_0$. When regarding the damping for overlapped levels we observe the same particle size and concentration dependence as we found for Born impurities with prefactor enhancement for light impurity atoms, an extra crossover to linear-in-$L$ size
dependence that occurs in the latter case.

The T-matrix approach allows  to investigate the bound state of a phonon localized on the impurity. The wave functions of phonons localized in the depth of a particle rapidly decay with distance and barely feel the particle boundary. Neglecting all boundary effects we analyze the bound states within the bulk approach (or in the overlapped regime). For light impurities the bound states just above $\omega_0$ appear starting from some threshold value $U_{min}<0$ of the impurity potential,  whereas heavy atoms may create a bound state only inside the gap between optical and acoustic phonon branches. When the impurity potential $U \to U_{min}$, the phenomenon of resonant impurity scattering arises and drastically enhances the long wavelength phonon damping.

Now let us discuss how separated and overlapped regimes crosses over. Since the linewidth depends (among other things) on the quantum number
of a level (both directly and via the particle shape), various parts of the spectrum may simultaneously exist in different regimes. Information about these dependence could be extracted either from analytical calculations or (for nontrivial particle shapes) from numerical simulations. In order to identify the regime  of broadening for any couple of spectral lines one should compare the interlevel distance
and the sum of halfwidths for these lines, i.e.
 \begin{equation}\label{cross1}
 \omega_{n+1}   - \omega_n  = \frac{1}{2} \, \left[\, \Gamma_{n+1}  + \Gamma_n  \, \right].
\end{equation}
Notice that $\Gamma_n$ and $\Gamma_{n+1}$ could be in  different regimes if say the levels $\omega_{n-1}$ and $\omega_n$ are already overlapped while $\omega_{n}$ and $\omega_{n+1}$ are not.

Next, since the linewidths in different regimes reveal different size and disorder dependence, it is instructive to determine the characteristic size for these two quantities to be equal to each other. Roughly, dropping the coefficients one obtains:
\begin{equation}\label{cross2}
{\cal L}_{c} \sim \frac{a_0}{S}.
\end{equation}
Remarkably, the linewidths  (in both regimes) reaches the interlevel distance on the same scale. In different terms, ${\cal L}_{c}$ is the spatial scale for ballistic Thouless time~\cite{edwards1972} (the time for phonon to fly ballistically through the particle) and the phonon lifetime $1/\Gamma_n$ to coincide. Moreover, Eq.~\eqref{cross2} can be inverted, thus determining the critical disorder strength for a given particle size:
\begin{equation}\label{cross3}
  {\cal S}_{c} \sim \frac{a_0}{L}.
\end{equation}


\begin{acknowledgments}

The authors are thankful to Igor Gornyi for valuable comments.
This work is supported by the Russian Science Foundation (Grant No. 19-72-00031).

\end{acknowledgments}

\appendix*

\section{Disorder in EKFG formalism}
\label{SAppend}

The consideration given in the main body of this paper is presented in maximally general form applicable for both theories (DMM-BPM, EKFG) we developed. It provides the limitations on final results including the coefficients and some relevant dependencies. In this Appendix we illustrate how the problem can be solved to the end for some particular case. As an example we choose the EKFG theory of a cubic particle which could be treated analytically. Since the overlapped regime is a sort of trivial we concentrate on the regime of separated levels.

\subsection{Hamiltonian and quantization}

We start from the Hamiltonian of EKFG model
\begin{equation}\label{Ham1}
  \mathcal{H}_0 = \frac{1}{2} \int_\Omega d^3 \mathbf{r} \left[ \Pi^2 - C_1 (\nabla \Phi)^2 + C_2 \Phi^2 \right].
\end{equation}
Here quantized field in Schrodinger representation
\begin{equation}\label{Phi1}
  \Phi(\mathbf{r}) = \sum_n \frac{1}{\sqrt{2 \omega_n}} \left( b_n Y_n(\mathbf{r}) + b^+_n Y^*_n(\mathbf{r}) \right)
\end{equation}
and momentum operator
\begin{equation}\label{Pi1}
  \Pi(\mathbf{r})=  - \sum_n i \sqrt{\frac{\omega_n}{2}} \left( b_n Y_n(\mathbf{r}) - b^+_n Y^*_n(\mathbf{r}) \right)
\end{equation}
are expressed via the phonon creation-annihilation operators $b^+_n \, (b_n)$ and the eigenfunctions of EKFG model $Y_n$. They obey conventional commutation relation:
\begin{equation}\label{Comm1}
  [\Phi(\mathbf{r}),\Pi(\mathbf{r}^\prime)] = -i \delta(\mathbf{r} - \mathbf{r}^\prime).
\end{equation}
It is easy to show that the quantized Hamiltonian reads:
\begin{equation}\label{Ham2}
  \mathcal{H}_0 = \sum_n \omega_n (b^+_n b_n +1/2).
\end{equation}
We can define the causal phonon Green's function
\begin{equation}\label{Green1}
  D(\mathbf{r}_1,t_1,\mathbf{r}_2,t_2) = - i \langle 0| \hat{\text{T}}  \left( \Phi(\mathbf{r}_1,t_1) \Phi(\mathbf{r}_2,t_2) \right)  |0 \rangle.
\end{equation}
After simple calculations and Fourier transform we get:
\begin{equation}\label{Green2}
  D_0(\mathbf{r}_1,\mathbf{r}_2,\omega) = \sum_n \frac{Y_n(\mathbf{r}_1) Y_n(\mathbf{r}_2) }{\omega^2 - \omega^2_n + i 0},
\end{equation}
where the eigenfunctions $Y_n$ are real.

Considering the solution of Eq.~\eqref{EKFG} for a given frequency $\omega$ we obtain:
\begin{equation} \label{KFG1}
  (\omega^2-C_1)Y-C_2\triangle Y=0.
\end{equation}
First, we should solve the eigenproblem
\begin{equation} \label{Laplace}
  \triangle Y + q^2 Y=0, \quad Y|_{\partial \Omega}=0,
\end{equation}
where $q$ will be referred to as the momentum. Thus, the spectrum of the problem has the form:
\begin{equation}\label{Spec1}
  \omega^2_{\bf q}=C_2-C_1 q^2.
\end{equation}
This is the equivalent of Eq.~\eqref{omega 0}:
\begin{equation}\label{Spec2}
  \omega_{\bf q} = \omega_0 - \alpha q^2,
\end{equation}
where we put $\omega_0 = \sqrt{C_2}$ and $\alpha = C_1/(2 \sqrt{C_2}) $.

As an example  we consider the cubic nanoparticle with the edge $b$. In that case EKFG has obvious solution normalized to unity:
\begin{equation}
  \label{WFcub}
  Y_{\bf n}= \left( \frac{2}{b} \right)^{3/2} \!\!\! e^{-i \omega t} \sin{\frac{\pi n_x x}{b}} \sin{\frac{\pi n_y y}{b}} \sin{\frac{\pi n_z z}{b}},
\end{equation}
where vector ${\bf n}=(n_x,n_y,n_3)$ enumerates the eigenstates, and the eigenvalues are
\begin{equation}
  \omega_\mathbf{n}=\omega_0 - \alpha  \left(\frac{\pi}{b} \right)^2 (n^2_x+n^2_y+n^2_z).
\end{equation}

\subsection{Point-like impurities}

Variation of atom masses in lattice model corresponds to variation of parameters $C_{1,2}$ of Eq.~\eqref{Ham1} in the continuous model. In the range of EKFG approach validity $q \, a_0 \ll~1$ we can neglect the variation of $C_1$ . Thus, the disorder-induced perturbation of the Hamiltonian reads:
\begin{equation}\label{Pert1}
  \mathcal{\delta H} = \frac{1}{2} \int_\Omega d^3 \mathbf{r} \, \left[ C_2(\mathbf{r}) - \overline{C_2} \right] \, \Phi^2.
\end{equation}
Here $\overline{C_2} = \omega^2_0$ stands for the average value of $C_2(\mathbf{r})$. Evidently, $C_2(\mathbf{r}) - \overline{C_2}\approx - \omega^2_0 \delta m(\mathbf{r})/m $, and we can write:
\begin{eqnarray}\label{Pert2}
  \mathcal{\delta H} &=& - \frac{1}{4}\sum_{n,n^\prime} \int_\Omega d^3 \mathbf{r} \, \frac{\omega^2_0}{\sqrt{\omega_n \omega_{n^\prime}}} \frac{\delta m(\mathbf{r})}{m} \\ &\times& (b_n + b^+_n)(b_{n^\prime}+ b^+_{n^\prime}) Y_n(\mathbf{r}) Y_{n^\prime}(\mathbf{r}). \nonumber
\end{eqnarray}
Within the EKFG approximation the disorder correlation function for weak point-like impurities is given by
\begin{equation}\label{Corr1}
  \frac{ \langle \, \delta m(\mathbf{r}_1) \, \delta m(\mathbf{r}_2) \, \rangle}{m^2} = S \, V_0 \, \delta(\mathbf{r}_1-\mathbf{r}_2),
\end{equation}
where $V_0$ is the unit cell volume.

In the limit of weak impurities only the second diagram in Fig.~\ref{DiagScat}b is relevant. By virtue of Eq.~\eqref{Corr1} the correction to the phonon Green's function yields:
\begin{eqnarray} \label{GreenBorn1}
  \delta D(\mathbf{r}_1,\mathbf{r}_2,\omega) &=&  \frac{S V_0 \omega^4_0}{4} \int d^3\mathbf{r} \Biggl[  \sum_{n} \frac{Y_{n}(\mathbf{r}_1) Y_{n}(\mathbf{r}) }{\omega^2 - \omega^2_{n} + i 0} \\
  &\times&  \sum_{n^\prime} \frac{Y_{n^\prime}(\mathbf{r}) Y_{n^\prime}(\mathbf{r}) }{\omega^2 - \omega^2_{n^\prime} + i 0} \sum_{n^{\prime\prime}} \frac{Y_{n^{\prime\prime}}(\mathbf{r}) Y_{n^{\prime\prime}}(\mathbf{r}_2) }{\omega^2 - \omega^2_{n^{\prime\prime}} + i 0} \Biggr]. \nonumber
\end{eqnarray}
For further analysis it is useful to perform the discrete Fourier transform of equation
\begin{equation}\label{Dyson1}
   D(\mathbf{r}_1,\mathbf{r}_2,\omega) =  D_0(\mathbf{r}_1,\mathbf{r}_2,\omega) + \delta D(\mathbf{r}_1,\mathbf{r}_2,\omega).
\end{equation}
In order to execute this transformation we multiply Eq.~(\ref{Dyson1}) by $Y_{n}(\mathbf{r}_1) Y_{n}(\mathbf{r}_2)$ and integrate it over $\mathbf{r}_1$ and $\mathbf{r}_2$. This procedure eliminates the sum in Eq.~\eqref{Green2} and sums over $n$ and $n^{\prime\prime}$ in Eq.~\eqref{GreenBorn1} due to orthogonality of  eigenfunctions $Y_{n}$. We can rewrite Eq.~\eqref{Dyson1} introducing renormalized complex eigenfrequencies $\tilde{\omega}_n$ as follows:
\begin{equation}\label{Dyson2}
  \omega^2-\tilde{\omega}^2_n = \omega^2 - \omega^2_n - \Pi_n(\omega),
\end{equation}
where
\begin{equation}\label{SigmaW1}
  \Pi_n(\omega) = \frac{S \, V_0 \, \omega^4_0}{4} \int d^3\mathbf{r} \sum_{n^\prime} \frac{Y^2_{n}(\mathbf{r}) Y^2_{n^\prime}(\mathbf{r}) }{\omega^2 - \omega^2_{n^\prime} + i 0}
\end{equation}


Using eigenfunctions~\eqref{WFcub} we calculate the self-energy~\eqref{SigmaW1}:
\begin{equation}\label{SigmaW2}
  \Pi_n(\omega) = \frac{S \, V_0 \, \omega^4_0}{32 \, V} \sum_{\mathbf{n}^\prime} \frac{\prod_{i=x,y,z} (2 + \delta_{n_i n^\prime_i})}{\omega^2 - \omega^2_{\mathbf{n}^\prime} + i 0},
\end{equation}
where $V$ is the particle volume. We find the phonon damping analytically using self-consistent approach and isolated levels approximation:
\begin{equation}\label{Dyson3}
  \Pi_\mathbf{n}(\omega) =\frac{27 \, S \, V_0  \, \omega^4_0}{32 \, V} \frac{1}{\omega^2 - \omega^2_{\mathbf{n}}-\Pi_\mathbf{n}(\omega)},
\end{equation}
Solving this equation, we obtain:
\begin{equation}\label{SigmaW3}
  \Pi_\mathbf{n}(\omega) = \frac{(\omega^2 - \omega^2_{\mathbf{n}}) - \sqrt{(\omega^2 - \omega^2_{\mathbf{n}})^2 -4 \frac{27 S \omega^4_0}{32 N} }}{2}.
\end{equation}
The phonon line obeys the semi-circle law (see Fig.~\ref{SeparLine}), and the width of the phonon line is given by:
\begin{equation}\label{GammaW1}
  \Gamma_\mathbf{n} = \omega_0 \sqrt{\frac{27 S }{32 N}}.
\end{equation}

 For degenerate level $\mathbf{n}$ only the coefficient in Eq.~\eqref{GammaW1} should be modified. If two quantum numbers in $\mathbf{n}$ are equal to each other then the level is three-fold degenerate and instead of $27/32$ we get $(27+24)/32=51/32$. For six-fold degenerate level with all three quantum numbers different the multiplier is $79/32$.

Furthermore, from Eq.~\eqref{SigmaW2} one can find the real part of the correction. This contribution is proportional to~$S$.

Finally, from Eq.~\eqref{GammaW1} we precisely determine the characteristic crossover scale ${\cal L}_c$ for the first level:
\begin{equation}\label{Scale1}
  {\cal L}_{c} =  \,15  \pi^4 F^2 \, \frac{a_0}{S}.
\end{equation}
For a diamond $F \approx 0.008$, and prefactor in front of the model-free ratio in Eq.~\eqref{Scale1} is not too small ($\approx 0.093$).

\subsection{Smooth disorder}

Now we consider smooth Gaussian disorder with certain characteristic length scale $\sigma$. The correlator reads:
\begin{equation}\label{Corr2}
  \frac{ \langle \delta m(\mathbf{r}_1) \delta m(\mathbf{r}_2) \rangle}{m^2} =  \frac{S}{\sigma^3 (2 \pi)^{3/2}}e^{-(\mathbf{r}_1-\mathbf{r}_2)^2/2 \sigma^2}.
\end{equation}
Using this correlator we obtain the following correction to the phonon Green's function:
\begin{eqnarray} \label{GreenS1}
  && \delta D(\mathbf{r}_1,\mathbf{r}_2,\omega) =  \frac{S \, \omega^4_0}{4} \int d^3\mathbf{r} d^3\mathbf{r}^\prime \Biggl[  \sum_{n} \frac{Y_{n}(\mathbf{r}_1) Y_{n}(\mathbf{r}) }{\omega^2 - \omega^2_{n} + i 0}  \nonumber \\
  &\times& \sum_{n^\prime} \frac{Y_{n^\prime}(\mathbf{r}) Y_{n^\prime}(\mathbf{r}^\prime) }{\omega^2 - \omega^2_{n^\prime} + i 0} \sum_{n^{\prime\prime}} \frac{Y_{n^{\prime\prime}}(\mathbf{r}^\prime) Y_{n^{\prime\prime}}(\mathbf{r}_2) }{\omega^2 - \omega^2_{n^{\prime\prime}} + i 0} \Biggr] \frac{e^{-(\mathbf{r}-\mathbf{r}^\prime)^2/2 \sigma^2}}{\sigma^3 (2 \pi)^{3/2}}. \nonumber
\end{eqnarray}
For non-degenerate level $\mathbf{n}$ and $q^2_\mathbf{n} \sigma^2 \ll 1 $ we have only small correction to damping:
\begin{equation}\label{GammaS2}
  \Gamma_\mathbf{n} = \omega_0 \, \sqrt{\frac{27  \, S}{32 \, N} \, \left( 1 - \frac{2}{3} q^2_\mathbf{n} \sigma^2 \right)}.
\end{equation}
For three-fold degenerate level the damping reads:
\begin{equation}\label{GammaS3}
  \Gamma_\mathbf{n} = \omega_0 \sqrt{\frac{\left( 51 - 42 \, \frac{\pi^2 n^2_x \sigma^2}{b^2}  - 76 \, \frac{\pi^2 n^2_y \sigma^2}{b^2} \right) S}{32 N} },
\end{equation}
and for six-fold one the result is as follows:
\begin{equation}\label{GammaS4}
  \Gamma_\mathbf{n} = \omega_0 \sqrt{\frac{\left( 79 - 66 \, q^2_\mathbf{n} \sigma^2 \right) \, S}{32 \, N} }.
\end{equation}
Thus, at $q^2_\mathbf{n} \sigma^2 \ll 1$ smooth random potential yields almost the same result as point-like impurities. For first modes with highest frequencies this condition is equivalent to $\sigma \ll L$, i.e., the length scale of the random potential must be much smaller than the particle size.





\bibliography{KFG}

\end{document}